\documentclass[reqno]{amsart}
\usepackage{amssymb,amsmath,amsthm}
\usepackage{eucal}
\usepackage{color}
\usepackage{epic}
\usepackage{eepic}
\usepackage{epsfig}
\usepackage{graphicx,caption}
\usepackage{amsfonts}
\usepackage{url}

\renewcommand{\theequation}{\arabic{equation}}

\def\m@th{\mathsurround=0pt}
\mathchardef\bracell="0365
\def\upbrall{$\m@th\bracell$}
\def\undertilde#1{\mathop{\vtop{\ialign{##\crcr
    $\hfil\displaystyle{#1}\hfil$\crcr
     \noalign
     {\kern1.5pt\nointerlineskip}
     \upbrall\crcr\noalign{\kern1pt
   }}}}\limits}
\def\theequation{\arabic{section}.\arabic{equation}}

\def\det{{\mathrm{Det}}}
\def\cL{{\mathcal L}}

\def\N{{\mathbb N}}
\def\Z{{\mathbb Z}}
\def\abs#1{{\mid #1 \mid}}
\def\max{{\text{max}}}
\def\s{{\bf s}}

\newcounter{NN}
\newtheorem{proposition}[NN]{Proposition}

\newtheorem{definition}[NN]{Definition}

\newcommand{\be}{\begin{equation}}
\newcommand{\ee}{\end{equation}}
\newcommand{\bea}{\begin{eqnarray}}
\newcommand{\eea}{\end{eqnarray}}
\newcommand{\bse}{\begin{subequations}}
\newcommand{\ese}{\end{subequations}}
\newcommand{\nn}{\nonumber}

\newcommand{\wt}{\widetilde}

\newcommand{\mbe}{{\boldsymbol e}}

\newcommand{\bpr}{\begin{prop}}
\newcommand{\epr}{\end{prop}}

\begin{document}
\begin{flushright}

\today \\
\end{flushright}

\title[A higher analogue of discrete-time Toda and a QQD scheme]{Higher analogues of the
discrete-time Toda equation and the quotient-difference algorithm}

\author{Paul E. Spicer}
\address{Katholieke Universiteit Leuven, Departement Wiskunde, Celestijnenlaan 200B
B-3001 Leuven, Belgium}
\email{pauldoesmaths@gmail.com}

\author{Frank W. Nijhoff}
\address{School of Mathematics, University of Leeds, Leeds LS2 9JT, U.K.}
\email{nijhoff@maths.leeds.ac.uk}

\author{Peter H. van der Kamp}
\address{Department of Mathematics and Statistics, La Trobe University, Victoria 3086, Australia}
\email{p.vanderkamp@latrobe.edu.au}

\begin{abstract}
The discrete-time Toda equation arises as a universal equation for the relevant Hankel determinants
associated with one-variable orthogonal polynomials through the mechanism of adjacency, which amounts to
the inclusion of shifted weight functions in the orthogonality condition. In this paper we extend this
mechanism to a new class of two-variable orthogonal polynomials where the variables are related via an
elliptic curve. This leads to a `Higher order Analogue of the Discrete-time Toda' (HADT) equation  for the associated Hankel determinants,
together with its Lax pair, which is derived from the relevant recurrence relations for the orthogonal
polynomials. In a similar way as the quotient-difference (QD) algorithm is related to the discrete-time Toda
equation, a novel quotient-quotient-difference (QQD) scheme is presented for the HADT equation.
We show that for both the HADT equation and the QQD scheme, there exists well-posed $s$-periodic initial
value problems, for almost all $\s\in\Z^2$. From the Lax-pairs we furthermore derive invariants for
corresponding reductions to dynamical mappings for some explicit examples.
\end{abstract}
\maketitle

\section{Introduction}
\setcounter{equation}{0}

The discrete-time Toda equation, i.e. the time-discretized version of the usual Toda chain,
which is given by the following partial difference equation (P$\Delta$E)
\be \label{eq:dttodai}
\tau_{l-1,m-1}\tau_{l+1,m+1}=\tau_{l+1,m-1}\tau_{l-1,m+1}-\tau_{l,m}^2,
\ee
plays an important role in many areas of mathematical physics. It is probably the first integrable fully discrete
equation that can be found in the literature: it appears, albeit in slightly different form, for the first time in the paper
by Frobenius \cite{Frob} as an identity for certain determinants used in the determination of rational
approximations of functions given by power series. The latter are nowadays known as Pad\'e approximants, which, in fact,
were introduced by Frobenius 10 years prior to Pad\'e's work, (cf. \cite{Gragg} for a more modern account).
The discrete-time Toda equation re-entered the modern era through the literature of integrable systems \cite{Hirota}.
Hirota in \cite{Hirota} introduced the equation as a natural time-discretization of the famous Toda chain
equation, generalizing the latter to a partial difference equation on the space-time lattice (here
the lattice sites are labelled by the discrete independent variables $(l,m)\in\mathbb{Z}^2$).
Eq. \eqref{eq:dttodai} exhibits the prominent integrability features, such as the existence of multi-soliton type solutions and
the existence of an underlying linear problem (Lax pair). This P$\Delta$E has also appeared in physics, namely as the
nonlinear equation governing the spin-spin correlation functions of the two-dimensional Ising model, cf. \cite{MCW,Perk}.
The connection we investigate and generalize in this paper, is the emergence of \eqref{eq:dttodai} in the theory of
formal orthogonal polynomials, where the equation \eqref{eq:dttodai}
is connected to the well known quotient-difference (QD) algorithm of Rutishauser, \cite{Rutis}, cf.\cite{PGR},
\begin{align}
e_{l,m+1} + q_{l+1,m+1} &= q_{l+1,m} + e_{l+1,m}, \notag \\
e_{l,m+1} q_{l,m+1} &= q_{l+1,m}e_{l,m}. \label{eq:qdi}
\end{align}

The theory of formal orthogonal polynomials is a subject central to modern numerical analysis, in which orthogonalities are investigated on their general properties
regarding the recurrence structure from a formal point of view  (i.e., without specifying particular classes of weight functions), with a sight on the
construction of numerical algorithms, rather that on the analytic properties (such as the moment problem, or the problem of the behaviour of the zeroes of the polynomial)
arising from the particular properties of the weight functions.
In this area of research general constructions such as those of vector Pad\'e approximants,
adjacent orthogonal families (where connected sequences of orthogonality functionals are postulated),
and associated convergence acceleration algorithms, and factorisation methods have been developed, cf. e.g.
\cite{Brez1,Draux}. In this context, the QD algorithm emerges as a prominent method to locate the zeroes
of analytic functions or to compute convergence factors for asymptotic expansions through formal power
series, using a finite-difference scheme and continued fractions, cf. \cite{henrici}.
It was pointed out in \cite{PGR} that both the QD algorithm and certain convergence algorithms, \cite{GRP},
are intimately related to integrable discrete systems.
In fact, the famous $\varepsilon$-algorithm of Wynn, \cite{Wynn}, surprisingly turns out to be identical to
the lattice potential Korteweg-de Vries (KdV) equation, a well-known exactly integrable P$\Delta$E,
thus allowing us to interpret this numerical algorithm as a symplectic dynamical system with a rich solution
structure. Similarly, the famous ``missing identity
of Frobenius'' appearing as the rhombus rule in the Pad\'e tables and governing the stability of the $\varepsilon$-algorithm, cf. \cite{Wynn2},
can be identified as an exactly solvable P$\Delta$E closely related to the Toda lattice and discretisations of the KdV equation.

\begin{figure}[h]
\setlength{\unitlength}{8mm}
\begin{center}
\begin{picture}(4,4)(0,0)
\matrixput(0,0)(1,0){5}(0,1){5}{\circle{.05}}
\multiput(0,4)(2,-2){3}{\circle*{.15}}
\matrixput(1,3)(1,0){2}(0,1){2}{\circle*{.15}}
\matrixput(2,0)(1,0){2}(0,1){2}{\circle*{.15}}
\thinlines
\dottedline{.10}(0,4)(4,0)(2,0)(2,4)(0,4)
\end{picture}
\end{center}
\caption{\label{stencil} The stencil of the HADT equation
(\ref{eq:sigma11})}
\end{figure}
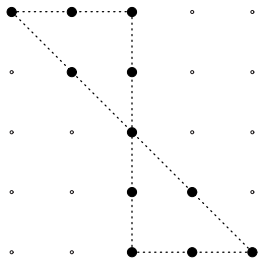

In this paper we introduce a novel integrable P$\Delta$E:
\def\Deltas{\sigma}
\bea
&&\Deltas_{l+1,m-2}\Deltas_{l-1,m+1}\left(\Deltas_{l,m+2}\Deltas_{l,m-1}-\Deltas_{l,m}\Deltas_{l,m+1}\right)
\nn \\
&=&\Deltas_{l,m-1}\Deltas_{l-1,m+2}\left(\Deltas_{l-1,m+1}\Deltas_{l+2,m-2}-\Deltas_{l,m}\Deltas_{l+1,m-1}\right)
\label{eq:11pt}\label{eq:sigma11} \\
&+&\Deltas_{l,m+1}\Deltas_{l+1,m-1}\left(\Deltas_{l,m-2}\Deltas_{l-1,m+2}
-\Deltas_{l+1,m-2}\Deltas_{l-2,m+2}\right),\ \nn
\eea
which we argue can be regarded as a Higher order Analogue of the Discrete-time Toda equation and therefore name it the HADT equation.
It is defined on a stencil of 11 points in the lattice as depicted in Figure \ref{stencil}.

The derivation of the HADT equation is parallel to the way in which the discrete-time Toda
equation emerges in the theory of orthogonal polynomials: in fact, we introduce a family of two-variable
orthogonal polynomials restricted by an elliptic curve and exploit the recurrence structure
for the relevant Hankel determinants. The general problem of orthogonal polynomials on algebraic curves
was discussed in the monograph by Suetin, \cite[Chapter 7]{Suetin}, and has also been addressed in the context of the study of formal
orthogonalities, cf. e.g. \cite{Brez3}. The construction of the associated two-variable orthogonal polynomials
on elliptic curve is pursued in \cite{Spicer}, whereas in this paper we will concentrate on the
HADT equation \eqref{eq:11pt} itself and its reductions. Nevertheless, we will present the derivation
from the adjacency structure of elliptic two-variable orthogonal polynomials, as it produces not only the
nonlinear equation itself but also its Lax pair. Furthermore,  the recurrence structure of the
relevant Hankel determinants also yields a novel type of QD formalism, namely the following so-called
quotient-quotient-difference (QQD) scheme:
\def\A{{u}}
\def\B{{v}}
\def\C{{w}}
\bea
\A_{l+2,m}+\B_{l+1,m}+\C_{l+1,m+1}&=&\A_{l,m+3}+\B_{l+1,m+1}+\C_{l+1,m}, \nn\\
\A_{l,m+3}\B_{l,m+1}&=&\B_{l+1,m}\A_{l+1,m}, \label{eq:qqdi}\\
\A_{l,m+3}\C_{l,m+1}&=&\C_{l+1,m+1}\A_{l+1,m+1}, \nn
\eea
which we believe may have future applications in numerical analysis.
\def\A{{U}}
\def\B{{V}}
\def\C{{W}}

Once we have obtained the HADT equation, the QQD scheme and their Lax pairs,
we address the issue of periodic reductions.
A periodic solution of a (system of) lattice equation(s) is a solution which satisfies
${\bf u}_{l,m}={\bf u}_{l+s_1,m+s_2}$ for some $s_1,s_2\in\Z$. By imposing periodicity
a lattice equation reduces to a system of ordinary difference equations,
or, equivalently, a mapping. The so-called staircase method utilizes the Lax-pair of
the lattice equation to construct integrals for the mapping. The method was
introduced in \cite{PNC}, where it was applied to the KdV
equation and to a mixed modified KdV-Toda equation, taking $s_1=s_2$ and
$s_1=s_2+1$. In \cite{QCPN} a two-parameter family of periodic reductions
was studied, with $s_1$ and $s_2$ being co-prime integers. Recently, in
\cite{K} a unified and geometric picture for periodic reductions, with nonzero
$\s=(s_1,s_2)\in\Z^2$ has been provided.
In \cite{K} it was shown, for any given scalar lattice equation on some
arbitrary stencil of lattice points, there exists a well-posed, or nearly
well-posed, $\s$-periodic initial value problem, for all nonzero $\s\in\Z^2$.
Therefore, the trace of the monodromy matrix provides integrals for any periodic reduction
of any integrable scalar lattice equation, and it is expected that the same is
true for systems of lattice equations, see \cite{K,KQ}. In this paper we generalize
the approach of \cite{K} and develop a systematic method to construct well-posed periodic
initial value problems for systems of lattice equations. This method is applied to
the QQD scheme as well as to an intermediate system of P$\Delta$Es, cf. equation
(\ref{eq:HS2}).

On general principles, one expects periodic reductions of integrable lattice equations to lead to integrable maps in a
precise sense, namely to complete integrability in the sense of Liouville-Arnold
\cite{BRSZ,Ves}. This requires, in addition to the existence of
a sufficient number of integrals, some global properties as well the
existence of a symplectic structure which is preserved by the discrete dynamics, and with respect to which
the integrals are in involution. The latter are issues which we are not addressing in the current
paper, where we have developed the basic structures and classified the consistent reductions from the
lattice equation together with the explicit form of the integrals.

Furthermore, there is the  problem of counting the number of independent integrals to ensure that in principle a 1-1 map to the
relevant number of degrees of freedoms exist for the mappings concerned.
In fact, typically some reduced variables, i.e. suitable combinations of the variables on the vertices,
need to be introduced, leading to a dimensional reduction of the mappings.
This issue is addressed in the section 5. For the reductions we obtained from the HADT equation, as well as those
from the QQD scheme, we are able to find a sufficient number of independent integrals, that is,
equal to at least half the dimensionality (i.e. number of components)  of the
mapping. For the reductions obtained from the intermediate system (\ref{eq:HS2}) this,
however, is not the case. In \cite{KQ} a systematic method was presented, which exploits
symmetries of the (system of) lattice equation(s) to reduce the dimension of the mappings.
For all cases that were considered in \cite{KQ}, the staircase method provides a sufficient
number of integrals for the dimensionally reduced mappings to be in principle completely
integrable. We arrive at a similar conclusion, however, in certain cases the staircase method
provides a product of 2-integrals $JJ^\prime$, from which a missing integral can be obtained.\footnote{Recall,
a function $J$ is an $k$-integral, or $k$-symmetry, of a mapping if it is an integral, or symmetry, of the
$k^\text{th}$ power of that mapping \cite{HBQC}. If one has one $k$-integral, then one can construct $k$ of them,
or, even better, $k$ integrals. For example, it is easy to see that $J^{\prime\prime}=J$ implies that both $JJ^\prime$
and $J+J^\prime$ are integrals.}

\section{The discrete-time Toda equation and the QD algorithm from orthogonal polynomials}
\setcounter{equation}{0}
We present here the derivation of the bilinear discrete-time Toda equation using formal orthogonal polynomials.
Recurrence relations for adjacent one-variable orthogonal polynomials provide a Lax-pair for this equation,
as well as for the associated QD algorithm. The results presented here are known, see for
example \cite{PGR}, however they outline the main idea which we employ in the setting of two-variable
elliptic orthogonal polynomials in the next section.

Let $P_n=P_n(x)$ be a family of polynomials of a variable $x$ orthogonal with respect to a weight function $w(x)$ on
a curve $\Gamma$ in the complex plane. The weight function defines a linear functional $\mathcal{L}$ with respect
to which we can define the moments $c_n$ as
\be
c_n=\mathcal{L}(\mbe_n)=\int_\Gamma x^{n}w(x)dx\quad i=0,1,\ldots
\ee
where $\mbe_n$ are the monomials.
The $P_n$'s, which are assumed to be monic, satisfy a three term recurrence relation of the form
\be\label{eq:mRec}
xP_{n}=P_{n+1}+S_{n}P_{n}+R_{n}P_{n-1}
\ee
(where $S_{n}$ and $R_{n}$ are finite) and in order to insure their existence we assume that the corresponding
Hankel determinants are non-zero.

We introduce a family of adjacent orthogonal polynomials (which are orthogonal with respect to a shifted weight $x^{m}w(x)$) \cite{Orthog}:
\bse\begin{equation}\label{eq:adjP}
P_{n}^{(m)}(x)\equiv
\frac{1}{\Delta_{n-1}^{(m)}}\left|\begin{array}{cccc}
c_{m}           & \ldots   &  \ldots     &  c_{n+m}  \\
\vdots          &             &                &  \vdots  \\
c_{n+m-1}    &             &               &  c_{2n+m-1}  \\
1                  & \ldots   &  \ldots     &  x^{n}
\end{array}\right|,
\end{equation}
with the corresponding Hankel determinant:
\begin{equation}\label{eq:delhan}
\Delta_{n}^{(m)}(x)\equiv
\left|\begin{array}{cccc}
c_{m}           & \ldots   &  \ldots     &  c_{n+m}  \\
\vdots          &             &                &  \vdots  \\
\vdots          &             &                &  \vdots  \\
c_{n+m}      & \ldots   &  \ldots     &  c_{2n+m}
\end{array}\right|,
\end{equation}\ese
and the two row/column Sylvester identity\footnote{This identity has many different names including the Jacobi identity, Lewis Carroll's
identity and the window-pane identity, however we will just refer to it as the Sylvester identity.} \cite{Bressoud,Muir},
\begin{equation}\label{eq:Sylv}
\left|
\begin{picture}(49,32)
\end{picture}
\right|\times
\left|
\begin{color}{red}
\begin{picture}(49,32)
\put(3,-22){\rule{.2mm}{1.8cm}}\put(46,-22){\rule{.2mm}{1.8cm}}
\put(1,-19){\rule{1.7cm}{.2mm}}\put(1,26){\rule{1.7cm}{.2mm}}
\end{picture}
\end{color}
\right| =
\left|
\begin{color}{red}
\begin{picture}(49,32)
\put(3,-22){\rule{.2mm}{1.8cm}}\put(1,26){\rule{1.7cm}{.2mm}}
\end{picture}
\end{color}
\right|\times\left|
\begin{color}{red}
\begin{picture}(49,32)
\put(46,-22){\rule{.2mm}{1.8cm}}\put(1,-19){\rule{1.7cm}{.2mm}}
\end{picture}
\end{color}
\right|
-\left|
\begin{color}{red}
\begin{picture}(49,32)
\put(3,-22){\rule{.2mm}{1.8cm}}\put(1,-19){\rule{1.7cm}{.2mm}}
\end{picture}
\end{color}
\right|\times\left|
\begin{color}{red}
\begin{picture}(49,32)
\put(46,-22){\rule{.2mm}{1.8cm}}\put(1,26){\rule{1.7cm}{.2mm}}
\end{picture}
\end{color}
\right| .
\end{equation}
Applying two different forms of the Sylvester identity to the determinant for $P_{n}^{(m)}$ leads to the derivation of two $xP_{n}^{(m)}$ recurrence relations
\bse\label{eq:genLP}\begin{eqnarray}
P_{n+1}^{(m)} & = & xP_{n}^{(m+1)}-V_{n}^{(m)}P_{n}^{(m)},  \\
P_{n+1}^{(m)} & = & xP_{n}^{(m+2)}-W_{n}^{(m)}P_{n}^{(m+1)},
\end{eqnarray}\ese
with
\begin{equation} \label{eq:VW}
V_{n}^{(m)}  =  \frac{\Delta_{n}^{(m+1)}\Delta_{n-1}^{(m)}}{\Delta_{n-1}^{(m+1)}\Delta_{n}^{(m)}} ,\quad
W_{n}^{(m)}  =  \frac{\Delta_{n}^{(m+1)}\Delta_{n-1}^{(m+1)}}{\Delta_{n}^{(m)}\Delta_{n-1}^{(m+2)}} .
\end{equation}
The combination of these relations leads to the monic recurrence relation, of the form (\ref{eq:mRec}),
\[
x P_{n}^{(m)}= P_{n+1}^{(m)}+(V_{n}^{(m)}+V_{n-1}^{(m+1)}-W_{n}^{(m)})P_{n}^{(m)}+(V_{n-1}^{(m+1)}-W_{n-1}^{(m)})V_{n-1}^{(m)}P_{n-1}^{(m)}. \]
From this approach the coefficients $S_{n}$ and $R_{n}$ can be further simplified (in terms of the Hankel determinants):
\bse\begin{eqnarray}
S_{n} & = & \frac{\Delta_{n}^{(m+1)}\Delta_{n-1}^{(m)}}{\Delta_{n}^{(m)}\Delta_{n-1}^{(m+1)}}+\frac{\Delta_{n-2}^{(m+1)}\Delta_{n}^{(m)}}{\Delta_{n-1}^{(m+1)}\Delta_{n-1}^{(m)}}  \label{eq:Sn} ,\\
R_{n} & = & \frac{\Delta_{n}^{(m)}\Delta_{n-2}^{(m)}}{\Delta_{n-1}^{(m)}\Delta_{n-1}^{(m)}}  , \label{eq:Rn}
\end{eqnarray}\ese
where we have suppressed the $m$-dependence in the symbols $R_n$ and $S_n$.
We achieve this simplification by making use of a bilinear relation that exists between the Hankel determinants $\Delta_{n}^{(m)}$.  This bilinear relation is found by applying the Sylvester identity to $\Delta_{n}^{(m)}$:
\begin{equation}
\left|
\begin{color}{red}
\begin{picture}(59,32)
\put(6,-26){\rule{.2mm}{2cm}}\put(53,-26){\rule{.2mm}{2cm}}
\put(1,-21){\rule{2cm}{.2mm}}\put(1,26){\rule{2cm}{.2mm}}
\begin{color}{black}
\put(20,2){\Large{$\Delta_{n}^{(m)}$}}
\end{color}
\end{picture}
\end{color}
\right|
\Rightarrow  \Delta_{n}^{(m)}\Delta_{n-2}^{(m+2)}=\Delta_{n-1}^{(m+2)}\Delta_{n-1}^{(m)}-\Delta_{n-1}^{(m+1)}\Delta_{n-1}^{(m+1)}
\label{eq:standh}.
\end{equation}
We will refer to equation (\ref{eq:standh}) as {\bf the discrete-time Toda equation}. Indeed, substituting
$\Delta^{(m)}_n=\tau_{-2n-m-1,m-1}$ and taking $l=-2n-m$ as new independent variable yields the more standard
form of this equation (\ref{eq:dttodai}). The continuum limit leading to the usual Toda chain is most easily
seen starting from equation (\ref{eq:dttodai}), multiplying with $4pq$ and applying a point transformation
$\tau_{l,m} \mapsto \tau_{l,m} \alpha^{l^2} \beta^{lm}$ with
$$
\alpha=\sqrt{\frac{p^2-q^2}{4pq}}, \quad \beta=\sqrt{\frac{p-q}{p+q}}
$$
leads to
\begin{equation}
\label{eq:lp}
(p-q)^2 \tau_{l-1,m-1} \tau_{l+1,m+1}-(p+q)^2 \tau_{l-1,m+1}\tau_{l+1,m-1}+4pq \tau_{l,m}^2 =0,
\end{equation}
in which $p,q$ are lattice parameters (associated with shifts on the lattice in $l$- and $m$ directions
respectively). Performing a changes of variables, $(l,m) \mapsto (n=l+m,m)$ and taking a limit
$l\rightarrow\infty$, $m\rightarrow\infty$, $n$ fixed, $q-p=\epsilon\rightarrow 0$, $\epsilon m \rightarrow t$,
Taylor expansion yields
$
\tau_{l,m}=\tau_n(t_0+\epsilon m)\quad\Rightarrow\quad \tau_{l,m+1}=\tau_{n+1}+\epsilon \dot{\tau}_{n+1}+ \dots
$
and we get the semi-discrete bilinear equation:
$$
4p^2 \left( \tau_n \ddot{\tau}_n- \dot{\tau}^2_n\right)= \tau_{n+2} \tau_{n-2},
$$
which is related to the usual Toda chain equation by setting $\phi_n=\log (\tau_{n}/\tau_{n-2})$.

Thus the shadows of integrability already appear in the underlying structure of the standard theory of
orthogonal polynomials.

\subsection{A Lax pair for the QD algorithm and the discrete-time Toda equation} \label{slax}
We now view equation (\ref{eq:standh}) as an integrable lattice equation, as opposed to
an identity for Hankel determinants (\ref{eq:delhan}).
\def\Deltau{\tau}
To this end we write $\Delta^{(m)}_n=\Deltau_{n,m}$,
and similarly $V^{(m)}_n=v_{n,m}$, $W^{(m)}_n=w_{n,m}$. The variable $x$ will play the role of spectral parameter
and will be denoted $x=\lambda$.

The relations (\ref{eq:genLP}) constitute a Lax-pair for equation (\ref{eq:standh}).
Working with the fields $v,w$ we first derive a related quotient-difference system.
Let $\Psi_{n,m}=(P^{(m)}_n,P^{(m+1)}_n)$, then using the recurrence relations
we derive
\be
\Psi_{n+1,m}=L_{n,m}\Psi_{n,m}\quad\textrm{and}\quad \Psi_{n,m+1}=M_{n,m}\Psi_{n,m}, \nn
\ee
with
\bse
\begin{eqnarray}
L_{n,m}&=&\left( \begin{array}{cc}
-v_{n,m} & \lambda \\
-v_{n,m} & \lambda + w_{n,m} -v_{n,m+1}
\end{array} \right), \\
M_{n,m}&=&\left( \begin{array}{cc}
0 & 1 \\
-\lambda^{-1} v_{n,m} & 1+\lambda^{-1} w_{n,m}
\end{array} \right).
\end{eqnarray}
\ese
The compatibility of the two linear systems is then equivalent to the discrete Lax equation:
\begin{eqnarray*}
0 &=& L_{n,m+1} M_{n,m} -  M_{n+1,m} L_{n,m} \\
  &=& \left( \begin{array}{cc}
  0 & 0 \\
  -\lambda^{-1} v_{n,m} E & E + \lambda^{-1} (w_{n,m}(v_{n+1,m}+E)-v_{n,m+1} w_{n+1,m})
  \end{array} \right),
\end{eqnarray*}
where
\[
E=v_{n,m+2}-v_{n+1,m}+w_{n+1,m}-w_{n,m+1}.
\]
Thus $L,M$ provide a Lax pair for the quotient-difference scheme
\bse \label{eq:QD}
\begin{eqnarray}
v_{n,m+2} + w_{n+1,m} &=& v_{n+1,m} + w_{n,m+1}, \label{eq:QDa} \\
w_{n,m} v_{n+1,m} &=& v_{n,m+1} w_{n+1,m}. \label{eq:QDb}
\end{eqnarray}
\ese
This scheme is related to the (more standard) QD algorithm (\ref{eq:qdi}),
by $q_{n,m}=v_{n,m}$, $e_{n,m}=v_{n,m+1}-w_{n,m}$.

Upon substitution of (\ref{eq:VW}) into (\ref{eq:QDa}) this equation is a consequence of
(\ref{eq:standh}), whereas substitution of (\ref{eq:VW}) into (\ref{eq:QDb}) turns into an identity.
To find a {\em good} Lax-pair for equation (\ref{eq:standh}), one substitutes equations (\ref{eq:VW})
into the above Lax matrices, and one uses (\ref{eq:standh}) to simplify
 \[
e_{n,m} = v_{n,m+1} - w_{n,m} = \frac{\Deltau_{n-1,m+1} \Deltau_{n+1,m} }{\Deltau_{n-1,m+1} \Deltau_{n,m} }.
 \]
 Using the same simplification, equation (\ref{eq:QDa}) can be written
 \begin{eqnarray*}
 0&=&e_{n,m+1}+w_{n+1,m}-v_{n+1,m}\\
 &=&\frac{\Deltau_{n+1,m+1}}{ \Deltau_{n,m+2}\Deltau_{n,m+1} \Deltau_{n+1,m} }\left(
 \Deltau_{n+1,m}\Deltau_{n-1,m+2}-\Deltau_{n,m+2}\Deltau_{n,m} + (\Deltau_{n,m+1})^2 \right),
 \end{eqnarray*}
where one recognizes the discrete-time Toda equation (\ref{eq:standh}).

\section{The HADT equation and a QQD scheme from elliptic orthogonal polynomials}
\setcounter{equation}{0}

In this section we derive the HADT equation.
We follow a similar route as in the previous section, except that here we consider
two-variable orthogonal polynomials, where the variables are restricted
by the condition that they form the coordinates of an elliptic curve.
The recurrence relations for these elliptic orthogonal polynomials yield
both a Lax pair and the QQD scheme.


\subsection{Two-variable elliptic orthogonal polynomials}
As a starting point for our construction we introduce the sequence of elementary monomials \cite{Spicer}, associated with a class of two variable orthogonal polynomials. We consider a sequence of monomials where the $x$ and $y$ are given different weight, namely $x\sim2$ and $y\sim3$:
\[
\mbe_0=1,\quad
\mbe_2=x,\quad
\mbe_3=y,\quad
\mbe_4=x^2,\quad
\mbe_5=xy,\quad
\mbe_6=x^3,\quad \cdots \]
or, in general:
\[
\mbe_0(x,y)=1,\quad\mbe_{2k}(x,y)=x^k,\quad \mbe_{2k+1}(x,y)=x^{k-1}y,\quad k=1,2,\dots .
\]
In comparison with the polynomials of Krall and Scheffer \cite{Krall}, their two-variable orthogonal polynomials consist of $x\sim1$ and $y\sim1$ of equal weight. The reason for choosing different weights is the variables $x$ and $y$ are related through a Weierstrass elliptic curve.
We use this sequence as our basis for the expansion of a new class of two-variable orthogonal polynomials taking the form:
\begin{equation}\label{eq:Ellpols}
P_k(x,y)=\sum_{j=0}^k p_j^{(k)}\mbe_j(x,y)\  ,
\end{equation}
which are \textit{monic} if the leading coefficient $p^{(k)}_k=1$.

In the spirit of the formal approach to orthogonal polynomials, cf. e.g. \cite{Brez0,Brez1}, we assume that a bilinear form $\langle,\rangle$ exists and can be derived from a linear functional $\mathcal{L}$ and consequently we can define the associated \textit{moments} by
\begin{equation}\label{eq:moments}
c_k=\mathcal{L}(\mbe_k)\   .
\end{equation}
The assumption of the existence of an inner product ~$\langle,\rangle$~ on the space $\mathcal{V}$ spanned by the monomials $\mbe_k$, is such that
\[ \langle xP,Q\rangle=\langle P,xQ\rangle\qquad,  \]
for any two elements $P,Q\in\mathcal{V}$.

Under the assumption of orthogonality and using the standard Gram-Schmidt orthogonalisation (through the use of
Cramer's rule), leads to the following expression for the adjacent elliptic orthogonal polynomials:
\begin{equation}\label{eq:Ppol}
P^{(l)}_k(x,y)\equiv
\left|\begin{array}{ccccc}
\langle\mbe_l,\mbe_0\rangle &\langle\mbe_l,\mbe_2\rangle & \cdots &\cdots &\langle\mbe_l,\mbe_k\rangle \\
\langle\mbe_{l+1},\mbe_0\rangle &\langle\mbe_{l+1},\mbe_2\rangle & \cdots &\cdots &\langle\mbe_{l+1},\mbe_k\rangle \\
\vdots  & \vdots  &   &  & \vdots \\
\vdots  & \vdots  &   &  & \vdots \\
\langle\mbe_{l+k-2},\mbe_0\rangle &\langle\mbe_{l+k-2},\mbe_2\rangle & \cdots &\cdots &\langle\mbe_{l+k-2},\mbe_k\rangle \\
\mbe_0  &\mbe_2  & \cdots &\cdots &\mbe_k
\end{array}\right|\big/ \Delta^{(l)}_{k-1}\quad , \quad l\neq 0\quad ,
\end{equation}
together with the corresponding Hankel determinant:
\begin{equation}\label{eq:PHankel}
\Delta^{(l)}_k=\left|\begin{array}{ccccc}
\langle\mbe_l,\mbe_0\rangle &\langle\mbe_l,\mbe_2\rangle & \cdots &\cdots &\langle\mbe_l,\mbe_k\rangle \\
\langle\mbe_{l+1},\mbe_0\rangle &\langle\mbe_{l+1},\mbe_2\rangle & \cdots &\cdots &\langle\mbe_{l+1},\mbe_k\rangle \\
\vdots  & \vdots  &   &  & \vdots \\
\vdots  & \vdots  &   &  & \vdots \\
\langle\mbe_{l+k-1},\mbe_0\rangle &\langle\mbe_{l+k-1},\mbe_2\rangle & \cdots &\cdots &\langle\mbe_{l+k-1},\mbe_k\rangle
\end{array}\right|\quad ,\quad l\neq 0\quad ,
\end{equation}
where for $l=0$:
\begin{equation}\label{eq:0Ppol}
P^{(0)}_k(x,y)\equiv
\left|\begin{array}{ccccc}
\langle\mbe_0,\mbe_0\rangle &\langle\mbe_0,\mbe_2\rangle & \cdots &\cdots &\langle\mbe_0,\mbe_k\rangle \\
\langle\mbe_{2},\mbe_0\rangle &\langle\mbe_{2},\mbe_2\rangle & \cdots &\cdots &\langle\mbe_{2},\mbe_k\rangle \\
\vdots  & \vdots  &   &  & \vdots \\
\vdots  & \vdots  &   &  & \vdots \\
\langle\mbe_{k-1},\mbe_0\rangle &\langle\mbe_{k-1},\mbe_2\rangle & \cdots &\cdots &\langle\mbe_{k-1},\mbe_k\rangle \\
\mbe_0  &\mbe_2  & \cdots &\cdots &\mbe_k
\end{array}\right|\big/ \Delta^{(0)}_{k-1}\quad ,
\end{equation}
with
\begin{equation}\label{eq:0PHankel}
\Delta^{(0)}_k=\left|\begin{array}{ccccc}
\langle\mbe_0,\mbe_0\rangle &\langle\mbe_0,\mbe_2\rangle & \cdots &\cdots &\langle\mbe_0,\mbe_k\rangle \\
\langle\mbe_{2},\mbe_0\rangle &\langle\mbe_{2},\mbe_2\rangle & \cdots &\cdots &\langle\mbe_{2},\mbe_k\rangle \\
\vdots  & \vdots  &   &  & \vdots \\
\vdots  & \vdots  &   &  & \vdots \\
\langle\mbe_{k},\mbe_0\rangle &\langle\mbe_{k},\mbe_2\rangle & \cdots &\cdots &\langle\mbe_{k},\mbe_k\rangle
\end{array}\right|\quad ,
\end{equation}

\paragraph{} In addition to the polynomials (\ref{eq:Ppol}), we also need  to introduce the polynomials:
\begin{equation}\label{eq:Qpol}
Q^{(l)}_k(x,y)\equiv \left|\begin{array}{ccccc}
\langle\mbe_l,\mbe_0\rangle &\langle\mbe_l,\mbe_2\rangle & \cdots &\cdots &\langle\mbe_l,\mbe_k\rangle \\
\langle\mbe_{l+2},\mbe_0\rangle &\langle\mbe_{l+2},\mbe_2\rangle & \cdots &\cdots &\langle\mbe_{l+2},\mbe_k\rangle \\
\vdots  & \vdots  &   &  & \vdots \\
\vdots  & \vdots  &   &  & \vdots \\
\langle\mbe_{l+k-1},\mbe_0\rangle &\langle\mbe_{l+k-1},\mbe_2\rangle & \cdots &\cdots &\langle\mbe_{l+k-1},\mbe_k\rangle \\
\mbe_0  &\mbe_2  & \cdots &\cdots &\mbe_k
\end{array}\right|\big/ \Theta^{(l)}_{k-1}\quad ,
\end{equation}
together with its corresponding Hankel determinant:
\begin{equation}\label{eq:QHankel}
\Theta^{(l)}_k=\left|
\begin{array}{ccccc}
\langle\mbe_l,\mbe_0\rangle &\langle\mbe_l,\mbe_2\rangle & \cdots &\cdots &\langle\mbe_l,\mbe_k\rangle \\
\langle\mbe_{l+2},\mbe_0\rangle &\langle\mbe_{l+2},\mbe_2\rangle & \cdots &\cdots &\langle\mbe_{l+2},\mbe_k\rangle \\
\vdots  & \vdots  &   &  & \vdots \\
\vdots  & \vdots  &   &  & \vdots \\
\langle\mbe_{l+k},\mbe_0\rangle &\langle\mbe_{l+k},\mbe_2\rangle & \cdots &\cdots &\langle\mbe_{l+k},\mbe_k\rangle
\end{array}\right|\  ,
\end{equation}
noting that
\[ Q_k^{(0)}=P_k^{(0)}\qquad , \qquad \Theta_k^{(0)}=\Delta_k^{(0)}\   . \]

\paragraph{Remark:} We note that for $l\neq 0,1$ the polynomials $P_k^{(l)}$ do \textit{not} form
an orthogonal family. In fact, from the determinantal definition (\ref{eq:Ppol}) we immediately observe that
\[ \langle \mbe_l,P_k^{(l)}\rangle =\langle \mbe_{l+1},P_k^{(l)}\rangle=\cdots
=\langle \mbe_{l+k-2},P_k^{(l)}\rangle=0\quad ,\quad
\langle \mbe_{l+k-1},P_k^{(l)}\rangle=\frac{\Delta_k^{(l)}}{\Delta_{k-1}^{(l)}}\   ,  \]
whereas
\[ \langle \mbe_{l-1},P_k^{(l)}\rangle=(-1)^{k-1} \frac{\Delta_k^{(l-1)}}{\Delta_{k-1}^{(l)}}\   .  \]

We now proceed using determinantal identities of Sylvester type (Appendix A) to derive relations between the polynomials $P_{k}^{(l)}$ and
the Hankel determinants.

\subsection{Recurrence relations in $P_{k}^{(l)}$ and $Q_{k}^{(l)}$}

Using a 3 row/column Sylvester identity we find it is possible to find $x$-recurrence relations and linear relations in $P_{k}^{(l)}$ and $Q_{k}^{(l)}$. To achieve the former it is necessary to fix the columns, so that $\mbe_{0}=1$ and $\mbe_{3}=y$ are removed from the determinant and the position of the row removal is dependent on restricting the introduction of new objects.  Hence we apply the following cutting of three rows and columns (\ref{eq:3Sylv}) to the determinant for $P_k^{(l)}$:

{\tiny\begin{eqnarray*}
\left|
\begin{picture}(59,32)
\end{picture}
\right|&\times&
\left|
\begin{color}{red}
\begin{picture}(59,32)
\put(6,-26){\rule{.3mm}{2cm}}\put(9,-5){\rule{.2mm}{0.5cm}}\put(12,-26){\rule{.3mm}{2cm}}\put(53,-26){\rule{.3mm}{2cm}}
\put(1,-21){\rule{2cm}{.3mm}}\put(1,-16){\rule{2cm}{.3mm}}\put(1,26){\rule{2cm}{.3mm}}
\end{picture}
\end{color}
\right|
 =  \left|
\begin{color}{red}
\begin{picture}(59,32)
\put(6,-26){\rule{.3mm}{2cm}}\put(9,-5){\rule{.2mm}{0.5cm}}\put(12,-26){\rule{.3mm}{2cm}}\put(1,-16){\rule{2cm}{.3mm}}\put(1,26){\rule{2cm}{.3mm}}
\end{picture}
\end{color}
\right|\times\left|
\begin{color}{red}
\begin{picture}(59,32)
\put(53,-26){\rule{.3mm}{2cm}}\put(1,-21){\rule{2cm}{.3mm}}
\end{picture}
\end{color}
\right| \\
&-&\left|
\begin{color}{red}
\begin{picture}(59,32)
\put(6,-26){\rule{.3mm}{2cm}}\put(9,-5){\rule{.2mm}{0.5cm}}\put(12,-26){\rule{.3mm}{2cm}}\put(1,-21){\rule{2cm}{.3mm}}\put(1,26){\rule{2cm}{.3mm}}
\end{picture}
\end{color}
\right|\times\left|
\begin{color}{red}
\begin{picture}(59,32)
\put(53,-26){\rule{.3mm}{2cm}}\put(1,-16){\rule{2cm}{.3mm}}
\end{picture}
\end{color}
\right|
 +  \left|
\begin{color}{red}
\begin{picture}(59,32)
\put(6,-26){\rule{.3mm}{2cm}}\put(9,-5){\rule{.2mm}{0.5cm}}\put(12,-26){\rule{.3mm}{2cm}}\put(1,-16){\rule{2cm}{.3mm}}\put(1,-21){\rule{2cm}{.3mm}}
\end{picture}
\end{color}
\right|\times\left|
\begin{color}{red}
\begin{picture}(59,32)
\put(1,26){\rule{2cm}{.3mm}}\put(53,-26){\rule{.3mm}{2cm}}
\end{picture}
\end{color}
\right|
\end{eqnarray*}}
(where the small red line indicates a space between the first column and the third). We are led to the recurrence relation
\begin{equation}\label{eq:Precurr}
P_k^{(l)}=x P_{k-2}^{(l+3)} - \B^{(l)}_{k-2} P_{k-1}^{(l)}
+\C^{(l)}_{k-2} P_{k-1}^{(l+1)},
\quad l\neq 0,1,
\end{equation}
where
\begin{equation}\label{eq:BC}
\B^{(l)}_k=\frac{\Delta_k^{(l)} \Delta^{(l+3)}_k}{\Delta_{k+1}^{(l)}\Delta_{k-1}^{(l+3)}},\qquad
\C^{(l)}_k=\frac{\Delta_k^{(l+1)} \Delta^{(l+2)}_k}{\Delta_{k+1}^{(l)}\Delta_{k-1}^{(l+3)}}.
\end{equation}
For $l=0$ we have
\begin{equation}\label{eq:0Precurr}
P_k^{(0)}=x P_{k-2}^{(4)} -~\frac{\Delta_{k-2}^{(0)} \Delta^{(4)}_{k-2}}{\Delta_{k-1}^{(0)}\Delta_{k-3}^{(4)}}~P_{k-1}^{(0)}
+~\frac{\Delta_{k-2}^{(2)}\Theta^{(2)}_{k-2}}{\Delta_{k-1}^{(0)}\Delta_{k-3}^{(4)}}~P_{k-1}^{(2)},
\end{equation}
whilst obviously, since $P_k^{(1)}$ is not defined, there is no relation for $l=1$.

By making use of {\it intermediate} determinant expressions (Appendix \ref{appendix2}) we can derive additional relations in terms of $P_{k}^{(l)}$ and $Q_{k}^{(l)}$.
Thus in addition to the recurrence relation (\ref{eq:Precurr}) we have also derived the following relation in terms of $P_{k}^{(l)}$ (equation \ref{eq:Tpol}):
\begin{equation}\label{eq:PPrel}
P_k^{(l)}= P_k^{(l+1)} +\A^{(l)}_{k-1}  P_{k-1}^{(l+1)},
\end{equation}
with
\begin{equation}\label{eq:A}
\A^{(l)}_k=\frac{\Delta^{(l)}_{k+1}\Delta_{k-1}^{(l+1)}}{\Delta^{(l)}_{k}\Delta_{k}^{(l+1)}} .
\end{equation}
Similarly, we derived recurrence formulae for the set $\{P_k^{(l)},Q_k^{(l)}\}$ (Appendix \ref{appendix2c}, equations (\ref{eq:xQrec}),(\ref{eq:QPPrel}) and (\ref{eq:QPQrel}) respectively),
\begin{eqnarray}
P_k^{(l)}&=&x Q_{k-2}^{(l+2)} -~\frac{\Delta_{k-2}^{(l)} \Theta^{(l+2)}_{k-2}}{\Delta_{k-1}^{(l)}\Theta_{k-3}^{(l+2)}}~P_{k-1}^{(l)}
+~\frac{\Theta_{k-2}^{(l)} \Delta^{(l+2)}_{k-2}}{\Delta_{k-1}^{(l)}\Theta_{k-3}^{(l+2)}}~Q_{k-1}^{(l)},
\quad l\neq 0,1\ \nn \\
&& \\
Q_k^{(l)}&=& P_k^{(l+1)} + \frac{\Delta^{(l)}_k\Delta_{k-2}^{(l+2)}}{\Theta^{(l)}_{k-1}\Delta_{k-1}^{(l+1)}} P_{k-1}^{(l+2)}, \\
Q_k^{(l)}&=&P_k^{(l)}-\frac{\Delta_k^{(l)}\Theta_{k-2}^{(l)}}{\Delta_{k-1}^{(l)}\Theta_{k-1}^{(l)}} Q_{k-1}^{(l)}.
\end{eqnarray}

\subsection{Bilinear Hankel identities}
In appendix \ref{appendix2b} we construct a pair of bilinear relations: a four term bilinear relation (derived by applying a 3 row/column Sylvester identity to the determinant $\Delta_{k}^{(l)}$), and a three term bilinear relation (resulting from the combination of other Hankel identities):
\bse\label{eq:HS}\begin{eqnarray}
\left|
\begin{color}{red}
\begin{picture}(59,32)
\put(6,-26){\rule{.3mm}{2cm}}\put(9,-5){\rule{.2mm}{0.5cm}}\put(12,-26){\rule{.3mm}{2cm}}\put(53,-26){\rule{.3mm}{2cm}}
\put(1,-21){\rule{2cm}{.3mm}}\put(1,21){\rule{2cm}{.3mm}}\put(1,26){\rule{2cm}{.3mm}}
\end{picture}
\end{color}
\right|
\Rightarrow
\Delta_k^{(l)}\Delta_{k-3}^{(l+4)}&=&\Delta_{k-1}^{(l)}\Delta_{k-2}^{(l+4)}-\Theta_{k-1}^{(l)}\Delta_{k-2}^{(l+3)}+\Delta_{k-1}^{(l+1)}\Theta_{k-2}^{(l+2)}, \nn \\
\label{eq:Hankrel}  \\
\Delta_{k}^{(l)}\Delta_{k-1}^{(l+2)}&=&\Theta_k^{(l)}\Delta_{k-1}^{(l+1)}-\Theta_{k-1}^{(l)}\Delta_{k}^{(l+1)}. \label{eq:Hankrel7}
\end{eqnarray}\ese

\subsection{The HADT equation}
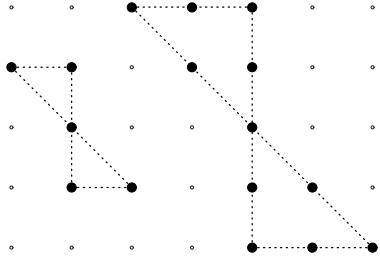
\begin{figure}[h]
\setlength{\unitlength}{8mm}
\begin{center}
\begin{picture}(6,4)(0,0)
\matrixput(0,0)(1,0){7}(0,1){5}{\circle{.05}}
\multiput(0,3)(2,-2){2}{\circle*{.15}}
\multiput(1,1)(0,1){3}{\circle*{.15}}
\multiput(2,4)(2,-2){3}{\circle*{.15}}
\matrixput(3,3)(1,0){2}(0,1){2}{\circle*{.15}}
\matrixput(4,0)(1,0){2}(0,1){2}{\circle*{.15}}
\thinlines
\dottedline{.10}(0,3)(2,1)(1,1)(1,3)(0,3)
\dottedline{.10}(2,4)(6,0)(4,0)(4,4)(2,4)
\end{picture}
\end{center}
\caption{\label{sof} The stencils of the discrete-time Toda equation (\ref{eq:standh}) and the HADT equation
(\ref{eq:quadd})}
\end{figure}

The system (\ref{eq:HS}) can be rearranged to eliminate the $\Theta_{k}^{(l)}$.  Thus we have
\begin{eqnarray*}
\frac{\Delta_{k}^{(l)}\Delta_{k-3}^{(l+4)}}{\Delta_{k-1}^{(l+1)}\Delta_{k-2}^{(l+3)}}-\frac{\Delta_{k-1}^{(l)}\Delta_{k-2}^{(l+4)}}{\Delta_{k-1}^{(l+1)}\Delta_{k-2}^{(l+3)}} & = & \frac{\Theta_{k-2}^{(l+2)}}{\Delta_{k-2}^{(l+3)}}-\frac{\Theta_{k-1}^{(l)}}{\Delta_{k-1}^{(l+1)}}, \\
\frac{\Delta_{k}^{(l)}\Delta_{k-1}^{(l+2)}}{\Delta_{k}^{(l+1)}\Delta_{k-1}^{(l+1)}} & = & \frac{\Theta_{k}^{(l)}}{\Delta_{k}^{(l+1)}}-\frac{\Theta_{k-1}^{(l)}}{\Delta_{k-1}^{(l+1)}},
\end{eqnarray*}
which can be expressed in a simpler way using
\begin{eqnarray*}
X_{k}^{(l)} & = & \Gamma_{k}^{(l)}-\Gamma_{k-1}^{(l)}, \\
Y_{k}^{(l)} & = & \Gamma_{k-2}^{(l+2)}-\Gamma_{k-1}^{(l)},
\end{eqnarray*}
where $\Gamma_{k}^{(l)}=\Theta_{k}^{(l)}/\Delta_{k}^{(l+1)}$.
From these two expressions we have
\bea
Y_{k+3}^{(l)}+X_{k+2}^{(l)}+X_{k+1}^{(l)}&=&\Gamma_{k+1}^{(l+2)}-\Gamma_{k}^{(l)}\nn \\
X_{k+1}^{(l+2)}+Y_{k+2}^{(l)}+X_{k+1}^{(l)}&=&\Gamma_{k+1}^{(l+2)}-\Gamma_{k}^{(l)}\nn
\eea
and by rearranging, we find a quadrilinear relation in terms of $\Delta_{k}^{(l)}$ on an 11-point stencil, the HADT equation:
\bea
&&\Delta_{k+1}^{(l)}\left(\Delta_{k}^{(l+4)}\Delta_{k}^{(l+1)}\Delta_{k-1}^{(l+3)}-\Delta_{k}^{(l+2)}\Delta_{k}^{(l+3)}\Delta_{k-1}^{(l+3)}
+\Delta_{k-2}^{(l+4)}\Delta_{k}^{(l+3)}\Delta_{k+1}^{(l+1)}\right) \nn  \\
&=&\Delta_{k-1}^{(l+4)}\left(\Delta_{k+2}^{(l)}\Delta_{k}^{(l+1)}\Delta_{k-1}^{(l+3)}-\Delta_{k}^{(l+2)}\Delta_{k}^{(l+1)}\Delta_{k+1}^{(l+1)}
+\Delta_{k}^{(l)}\Delta_{k}^{(l+3)}\Delta_{k+1}^{(l+1)}\right)\ . \nn  \\
&&\label{eq:quadd}
\eea
The HADT equation bears its name as a Higher order Analogue of the Discrete-time Toda equation, due to the similarities between how both equations arise in the theory of formal orthogonal polynomials. We expect the HADT equation to also be integrable and derive a Lax pair for it in the next section.
Also note the similarity between the stencils of the two equations (\ref{eq:standh}) and (\ref{eq:quadd}), as displayed in Figure \ref{sof}.

It would be interesting to know how to enter lattice parameters (i.e., parameters
associated with the grid steps) into the HADT equation, as we did for the
discrete-time Toda equation, see equation (\ref{eq:lp}), and then to study its continuum limit. However, there
is a lot of freedom here to enter lattice parameters through point transformations, and this
should be properly investigated. We leave that issue to a future work.

\def\Thetar{\rho}
Next we follow a similar approach as in section \ref{slax} and derive Lax pairs for the several
Hankel identities viewed as integrable systems, using the recurrence relations
for the two-variable elliptic orthogonal polynomials.
As before the fields will carry two sub-indices; we denote $x=\lambda$, $\Delta^{(m)}_l=\Deltas_{l,m}$,
$\Theta^{(m)}_l=\Thetar_{l,m}$, $\A^{(l)}_k=u_{k,l}$, etc.
By not evaluating the coefficients in the recurrences (\ref{eq:PPrel}) and (\ref{eq:Precurr}), we will find a novel
quotient-quotient-difference scheme.

\subsection{A Lax pair for the system of lattice equations (\ref{eq:HS})}
We derive a Lax pair for the system of equations (\ref{eq:HS}) which in terms of $\Deltas,\Thetar$ reads
\bse \label{eq:HS2}
\bea
\Deltas_{l+1,m-2}\Deltas_{l-2,m+2}+\Thetar_{l,m-2}\Deltas_{l-1,m+1}&=&\Deltas_{l,m-2}\Deltas_{l-1,m+2}
+\Deltas_{l,m-1}\Thetar_{l-1,m}, \nn \\
\label{eq:HS2a}  \\
\Deltas_{l,m-1}\Deltas_{l-1,m+1}+\Thetar_{l-1,m-1}\Deltas_{l,m}&=&\Thetar_{l,m-1}\Deltas_{l-1,m}.
\label{eq:HS2b}
\eea
\ese
Taking $\Psi_{l,m}=(Q^{(m+1)}_l, P_{l+1}^{(m+1)}, Q_{l+1}^{(m)}, P_{l+2}^{(m)})$ we derive
$\Psi_{l+1,m}=L_{l,m}\Psi_{l,m}$, and $\Psi_{l+1,m}=M_{l,m}\Psi_{l,m}$, with
{\footnotesize\begin{align*}
L_{l,m}=&\left( \begin{array}{cccc}
\frac{-\Deltas_{l+1,m+1}\Thetar_{l-1,m+1}}{\Deltas_{l,m+1}\Thetar_{l,m+1}} & 1 & 0 & 0 \\
0 & \frac{-\Deltas_{l+2,m}\Deltas_{l,m+1}}{\Deltas_{l+1,m}\Deltas_{l+1,m+1}} & 0 & 1\\
0 & 0 & \frac{-\Deltas_{l+2,m}\Thetar_{l,m}}{\Deltas_{l+1,m}\Thetar_{l+1,m}} & 1\\
0 & \frac{-\Deltas_{l+1,m+2}\Deltas_{l,m+1}}{\Deltas_{l+3,m}\Deltas_{l+1,m}} & 0 &
\frac{\Deltas_{l+1,m+2}\Deltas_{l+1,m+1}-\Deltas_{l+1,m}\Deltas_{l+1,m+3}}{\Deltas_{l,m+3}\Deltas_{l+2,m}}
\end{array} \right) \\
&+ \lambda \left( \begin{array}{cccc}
0 & 0 & 0 & 0 \\
0 & 0 & 0 & 0 \\
0 & 0 & 0 & 0 \\
\frac{\Deltas_{l+1,m+2}\Thetar_{l-1,m+1}}{\Deltas_{l,m+1}\Deltas_{l,m+3}} &
1+\frac{\Thetar_{l+1,m+1}\Thetar_{l,m}}{\Deltas_{l,m+3}\Deltas_{l+1,m}} &
\frac{-\Thetar_{l+1,m+1}\Thetar_{l,m}}{\Deltas_{l,m+3}\Deltas_{l+1,m}} & 0
\end{array} \right),
\end{align*}}
and
\begin{align*}
M_{l,m}=&\left( \begin{array}{cccc}
\frac{-\Thetar_{l,m+2}\Deltas_{l,m+2}\Thetar_{l-1,m+1}}{\Thetar_{l-1,m+2}\Deltas_{l,m+1}\Deltas_{l,m+3}} &
\frac{-\Thetar_{l,m+2}\Thetar_{l,m+1}\Thetar_{l,m}}{\Thetar_{l-1,m+2}\Deltas_{l+1,m}\Deltas_{l,m+3}} &
\frac{\Thetar_{l,m+2}\Thetar_{l,m+1}\Thetar_{l,m}}{\Thetar_{l-1,m+2}\Deltas_{l+1,m}\Deltas_{l,m+3}} & 0 \\
0 & \frac{\Thetar_{l+1,m}\Deltas_{l,m+1}}{\Deltas_{l+1,m}\Deltas_{l,m+2}} &
\frac{-\Deltas_{l+1,m+1}\Thetar_{l,m}}{\Deltas_{l+1,m}\Deltas_{l,m+2}} & 0 \\
\frac{-\Deltas_{l+1,m+1}\Thetar_{l-1,m+1}}{\Deltas_{l,m+1}\Thetar_{l,m+1}} & 1 & 0 & 0 \\
0 & \frac{-\Deltas_{l+2,m}\Deltas_{l,m+1}}{\Deltas_{l+1,m}\Deltas_{l+1,m+1}} & 0 & 1
\end{array} \right) \\
&+ \lambda^{-1} \left( \begin{array}{cccc}
0 & \frac{\Deltas_{l,m+1}\Thetar_{l,m+2}\Deltas_{l,m+2}}{\Thetar_{l-1,m+2}\Deltas_{l+1,m}\Deltas_{l,m+3}} &
\frac{-\Thetar_{l,m}\Deltas_{l,m+2}}{\Thetar_{l-1,m+2}\Deltas_{l+1,m}} & 1-\frac{\Thetar_{l,m+2}\Deltas_{l-1,m+3}}{\Thetar_{l-1,m+2}\Deltas_{l,m+3}} \\
0 & 0 & 0 & 0 \\
0 & 0 & 0 & 0 \\
0 & 0 & 0 & 0
\end{array} \right).
\end{align*}
Here we have used the equations (\ref{eq:HS2}) to reduce the number of terms in the entries
of the Lax matrices as much as possible. The compatibility condition $L_{l,m+1} M_{l,m} -  M_{l+1,m} L_{l,m} = 0$ is
equivalent to the system (\ref{eq:HS2}). However, to verify that the left hand side vanishes modulo the system,
some work has to be done. The equations do not factor out as nicely as in section \ref{slax}.

\subsection{A Lax pair for the HADT equation, and the QQD scheme}
We first derive a Lax-pair for the following QQD scheme, satisfied
by the coefficients ($\A^{(m)}_l=u_{l,m}$ etc.) in the recurrence equations
(\ref{eq:Precurr}) and (\ref{eq:PPrel}):
\def\A{{u}}
\def\B{{v}}
\def\C{{w}}
\bse \label{eq:QQD}
\begin{align}
\A_{l+2,m}+\B_{l+1,m}+\C_{l+1,m+1}&=\A_{l,m+3}+\B_{l+1,m+1}+\C_{l+1,m}, \label{eq:QQDD}\\
\A_{l,m+3}\B_{l,m+1}&=\B_{l+1,m}\A_{l+1,m}, \label{eq:QQDQ1}\\
\A_{l,m+2}\C_{l,m}&=\C_{l+1,m}\A_{l+1,m}. \label{eq:QQDQ2}
\end{align}
\ese

Let $\Psi_{l,m}=(P^{(m+3)}_l, P^{(m+2)}_{l+1}, P^{(m+1)}_{l+2}, P^{(m)}_{l+3} )$. We use (\ref{eq:Precurr}) and (\ref{eq:PPrel}) to find
$\Psi_{l+1,m}=L_{l,m}\Psi_{l,m}$, and $\Psi_{l,m+1}=M_{l,m}\Psi_{l,m}$, where $L$ and $M$ are given
\begin{align*}
L_{l,m}&=\left( \begin{array}{cccc}
-\A_{l,m+2} & 1 & 0 & 0 \\
0 & -\A_{l+1,m+1} & 1 & 0 \\
0 & 0 & -\A_{l+2,m} & 1 \\
0 & 0 & -\C_{l+2,m}\A_{l+2,m} & -\B_{l+2,m}+\C_{l+2,m}
\end{array} \right) \\
&+ \lambda \left( \begin{array}{cccc}
0 & 0 & 0 & 0 \\
0 & 0 & 0 & 0 \\
0 & 0 & 0 & 0 \\
\A_{l+1,m+2}\A_{l,m+2} & -\A_{l+1,m+1}-\A_{l+1,m+2} & 1 & 0
\end{array} \right),
\end{align*}
and
\begin{align*}
M_{l,m}&=
\left( \begin{array}{cccc}
-\frac{\A_{l,m+2}}{\A_{l,m+3}} & \frac{1}{\A_{l,m+3}} & 0 & 0 \\
-\A_{l,m+2} & 1 & 0 & 0 \\
0 & -\A_{l+1,m+1} & 1 & 0 \\
0 & 0 & -\A_{l+2,m} & 1
\end{array} \right) \\
&+\lambda^{-1} \left( \begin{array}{cccc}
0 & -\C_{l,m+1} &  1+\frac{\C_{l+1,m}-\B_{l+1,m}}{\A_{l,m+3}} & -\frac{1}{\A_{l,m+3}} \\
0 & 0 & 0 & 0 \\
0 & 0 & 0 & 0 \\
0 & 0 & 0 & 0
\end{array} \right).
\end{align*}
Here we have used the scheme (\ref{eq:QQD}) to simplify the $\lambda^{-1}$-part of $M_{l,m}$.
The compatibility of these linear systems is equivalent to the QQD scheme (\ref{eq:QQD}). Note, by substituting the
expressions (\ref{eq:A}), (\ref{eq:BC}) for $\A,\B,\C$ in terms of $\Deltas$ into the QQD scheme, the first equation is equivalent to the HADT equation, whereas the later two are satisfied identically. Therefore, after the same substitution, the matrices $L,M$ provide a {\em good} Lax pair for the HADT equation (\ref{eq:quadd}). The stencils of the Lax matrices $L(\Delta),M(\Delta)$ are depicted in Figures \ref{SL}, \ref{SM}.

\noindent
\parbox{60mm}{
\begin{center}
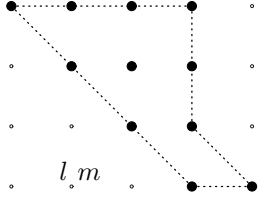

\setlength{\unitlength}{8mm}
\begin{picture}(4,3)(0,0)
\matrixput(0,0)(1,0){5}(0,1){4}{\circle{.05}}
\multiput(0,3)(1,0){4}{\circle*{.15}}
\multiput(1,2)(1,0){3}{\circle*{.15}}
\multiput(2,1)(1,0){2}{\circle*{.15}}
\multiput(3,0)(1,0){2}{\circle*{.15}}
\thinlines
\dottedline{.10}(0,3)(3,3)(3,1)(4,0)(3,0)(0,3)
\put(.8,.1){$l$}
\put(1.1,.1){$m$}
\end{picture}
\captionof{figure}{\label{SL} \\
The stencil of $L_{l,m}$}
\end{center}}
\hfill
\parbox{60mm}{
\setlength{\unitlength}{8mm}
\begin{center}
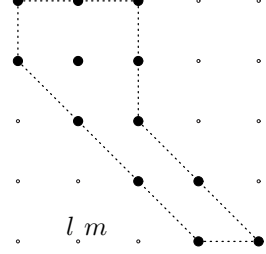

\begin{picture}(4,4)(0,0)
\matrixput(0,0)(1,0){5}(0,1){5}{\circle{.05}}
\multiput(0,4)(1,0){3}{\circle*{.15}}
\multiput(0,3)(1,0){3}{\circle*{.15}}
\multiput(1,2)(1,0){2}{\circle*{.15}}
\multiput(2,1)(1,0){2}{\circle*{.15}}
\multiput(3,0)(1,0){2}{\circle*{.15}}
\thinlines
\dottedline{.10}(0,4)(2,4)(2,2)(4,0)(3,0)(0,3)(0,4)
\put(.8,.1){$l$}
\put(1.1,.1){$m$}
\end{picture}
\captionof{figure}{\label{SM} \\
The stencil of $M_{l,m}$}
\end{center}
}
\section{Periodic reduction} \label{3}
\setcounter{equation}{0}
In \cite{K} a method was given to obtain initial value problems for scalar equations on arbitrary stencils.
The question was raised whether a similar construction can be done for systems, and one example was given
(the quotient-difference algorithm) where this is the case indeed. Here we will present a systematic approach
towards constructing initial value problems for systems of partial difference equations.

Using the method of \cite{K} it is easily shown that the dimension of the $\s$-periodic initial value problem for the HADT equation is given by the piecewise linear function
\be\label{pwlf}
4\max\{|s_1+s_2|,|s_1|\}.
\ee
We will show that the $\s$-periodic straight band initial value problems for the intermediate system (\ref{eq:HS2}) as well as those for the QQD scheme are of the same dimension as the initial value problems for the HADT equation, namely (\ref{pwlf}).

Although we do give a method to construct $\s$-periodic initial value problems for systems, the question remains whether this can be done algorithmically. As we
will see, one can not easily determine the dimensionality of the reductions, nor which regions for $\s$ should be distinguished, by simply looking at the stencils on which the equations of the system are defined.

\subsection{Initial value problems for the HADT equation} \label{ss1}

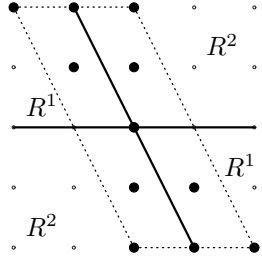
\begin{figure}[h]
\setlength{\unitlength}{8mm}
\begin{center}
\begin{picture}(4,4)(0,0)
\matrixput(0,0)(1,0){5}(0,1){5}{\circle{.05}}
\multiput(2,0)(0,1){5}{\circle*{.15}}
\multiput(0,4)(1,-1){5}{\circle*{.15}}
\multiput(1,4)(2,-4){2}{\circle*{.15}}
\thinlines
\dottedline{.10}(0,4)(2,0)(4,0)(2,4)(0,4)
\thicklines
\path(0,2)(4,2)
\path(1,4)(3,0)
\put(.2,.2){$R^2$}
\put(.2,2.2){$R^1$}
\put(3.5,1.2){$R^1$}
\put(3.2,3.2){$R^2$}
\end{picture}
\end{center}
\caption{\label{scr} The 11-point stencil of the HADT equation, its convex hull, and
two regions in the plane}
\end{figure}

To pose initial value problems for equation (\ref{eq:11pt}) we use the method
developed in \cite{K}. By the {\em $S$-polygon} of the equation we mean the boundary of the
convex hull of the {\em stencil $S$} of the equation (the dotted parallelogram in Figure \ref{scr}).
The {\em $S$-directions} are the directions in the $S$-polygon of the equation: $(1,0)$ and $(1,-2)$.
Therefore, we distinguish two regions in the plane
\begin{eqnarray*}
R^1&=&\{ \s \in \Z^2 : a\geq 0 , b>0 , \epsilon=1 \text{ or } b > 2a , \epsilon=-1 \}, \\
R^2&=&\{ \s \in \Z^2 : 0<b< 2a , \epsilon=-1 \},
\end{eqnarray*}
as in Figure \ref{scr}.
If $\s\in R^1$ or $\s\in R^2$, then performing an $\s$-periodic reduction yields a well-posed
initial value problem, i.e. the $\s$-periodic solution is obtained by iteration of a single valued mapping.
If on the other hand $\s$ is on a boundary between $R^1$ and $R^2$ one needs to impose periodicity on the solution,
and then one gets a multi-valued mapping, or correspondence.
To each of the regions $R^i$ we associate a vector $d^i$ which is a difference between the two points
of $S$, where a line with direction $\s\in R^i$ intersects the $S$-polygon in a single point.
For the HADT equation we have
\be \label{eq:diffs}
d^1=(0,4), \qquad d^2=(4,-4).
\ee
Hence the dimension of the $\s$-periodic initial value problem is, cf. \cite[Equation 8]{K},
\[
\max_{i=1,2}\{|\det \binom \s{d^i}|\}=4\max\{|s_1+s_2|,|s_1|\}.
\]
We give two examples before we present the general $\s$-periodic initial value problem in terms
of $\s$-reduced variables.

\def\z{{x}}
\subsubsection{(2,-1)-reduction of the HADT equation} \label{s2-1s}

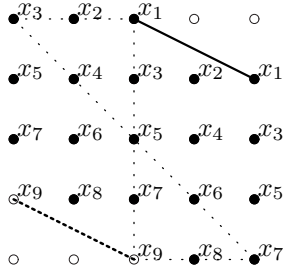
\begin{figure}[h]
\setlength{\unitlength}{8mm}
\begin{center}
\begin{picture}(4,4)(0,0)
\matrixput(0,0)(1,0){5}(0,1){5}{\circle{.15}}
\multiput(0,2)(0,1){3}{\circle*{.15}}
\matrixput(1,1)(1,0){2}(0,1){4}{\circle*{.15}}
\matrixput(3,0)(1,0){2}(0,1){4}{\circle*{.15}}

\dottedline{.2}(0,4)(2,4)(2,0)(4,0)(0,4)
\thicklines
\path(2,4)(4,3)
\dottedline{.12}(0,1)(2,0)

\multiput(0.06,4.06)(2,-1){3}{$\z_3$}
\multiput(0.06,3.06)(2,-1){3}{$\z_5$}
\multiput(0.06,2.06)(2,-1){3}{$\z_7$}
\multiput(0.06,1.06)(2,-1){2}{$\z_9$}
\multiput(1.06,4.06)(2,-1){2}{$\z_2$}
\multiput(1.06,3.06)(2,-1){2}{$\z_4$}
\multiput(1.06,2.06)(2,-1){2}{$\z_6$}
\multiput(1.06,1.06)(2,-1){2}{$\z_8$}
\multiput(2.06,4.06)(2,-1){2}{$\z_1$}
\end{picture}
\end{center}
\caption{\label{iv1eq} (2,-1)-periodic initial values}
\end{figure}

We specify initial values between two parallel lines that
squeeze the stencil as in Figure \ref{iv1eq}.
Clearly, given the 8 values $\z_1,\z_2,\ldots,\z_8$ at the black dots one is able to determine
the value $\z_9$ at the white dots on the fat-dotted line, using the HADT equation. Thus we get
an eight dimensional mapping:
\be \label{mq2-1}
\begin{split}
\z_i &\mapsto \z_{i+1}, \quad 1\leq i <8, \\
\z_8 &\mapsto \frac{\z_1\z_4\z_7\z_8}{\z_2\z_3\z_6}+\frac{\z_5\z_7}{\z_3}+\frac{\z_3\z_8}{\z_2} -\frac{\z_4\z_5\z_8}{\z_2\z_6}-\frac{\z_4\z_7^2}{\z_3\z_6}.
\end{split}
\ee

We can calculate integrals for this mapping by constructing a so called monodromy matrix, which
is a product of Lax matrices along a staircase over a one period long distance.
Taking the point at the bottom left corner to be the origin $(l,m)=(0,0)$, we define the {\em monodromy
matrix} to be the inversely ordered product of Lax matrices along a staircase from $(l,m)=(1,0)$ to
$(l,m)=(3,-1)$,
\[
\cL=M_{3,-1}^{-1}L_{2,0}L_{1,0}.
\]
Using the stencils of the Lax matrices, see Figures \ref{SL} and \ref{SM}, the reader can verify that
the matrices in $\cL$ depend on initial values only. The characteristic
polynomial of  $\cL$ is
\[
\det(\mu {\rm I} - \cL) = \mu^2((\mu -\lambda)^2 + 1) + \mu I_1(\mu^2-\lambda\mu-\lambda)
- \lambda(I_2\mu^2 + I_3 \mu + 1),
\]
where
\begin{eqnarray*}
I_1&=&\frac{\z_3\z_6}{\z_2\z_7}+\frac{\z_2\z_7}{\z_6\z_3}-\frac{\z_1\z_8}{\z_6\z_3}
+\frac{\z_1\z_4}{\z_2\z_3}+\frac{\z_8\z_5}{\z_6\z_7}-\frac{\z_5\z_4}{\z_2\z_7}, \\
I_2&=&\left(\frac{\z_5^2}{\z_4\z_6}+\frac{\z_3^2}{\z_2\z_4}+\frac{\z_1\z_7}{\z_2\z_6}
-\frac{\z_3\z_5}{\z_2\z_6}\right)
\left(\frac{\z_6^2}{\z_5\z_7}+\frac{\z_4^2}{\z_3\z_5}+\frac{\z_2\z_8}{\z_3\z_7}
-\frac{\z_4\z_6}{\z_3\z_7}\right), \\
I_3&=&\frac{\z_2\z_5}{\z_3\z_4}+\frac{\z_6\z_3}{\z_4\z_5}+\frac{\z_4\z_7}{\z_5\z_6}
+\frac{\z_1\z_8}{\z_3\z_6}-\frac{\z_2\z_7}{\z_3\z_6}.
\end{eqnarray*}
Because we subtracted $I_1$ from the coefficient of $ \mu^2\lambda$ in the definition of $I_2$
it factorizes nicely as $JJ^\prime$, where $J$ is a 2-integral of the map and $J^\prime$ is the
image of $J$ under the map. Remember a $k$-integral is an integral of the $k$th power of the map \cite{HBQC}.
Indeed, given that $J^{\prime\prime}=J$, we find that the product $JJ^\prime=J^{\prime\prime}J^\prime=(JJ^\prime)^\prime$ is an invariant. But now, we can write down
another (functionally independent) invariant, namely $J+J^{\prime}$. Moreover, the integrals
$I_1,I_2,I_3,I_4$, with
\[
I_4=\frac{\z_5^2}{\z_4\z_6}+\frac{\z_3^2}{\z_2\z_4}+\frac{\z_1\z_7}{\z_2\z_6}
-\frac{\z_3\z_5}{\z_2\z_6}+\frac{\z_6^2}{\z_5\z_7}+\frac{\z_4^2}{\z_3\z_5}+\frac{\z_2\z_8}{\z_3\z_7}
-\frac{\z_4\z_6}{\z_3\z_7},
\]
are functionally independent.

\subsubsection{(2,-3)-reduction of the HADT equation} \label{s2-3s}

\begin{figure}[h]
\setlength{\unitlength}{8mm}
\begin{center}
\begin{picture}(4,4)(0,0)
\matrixput(0,0)(1,0){5}(0,1){5}{\circle{.15}}
\multiput(1,2)(1,1){2}{\circle*{.15}}
\matrixput(0,3)(1,0){2}(0,1){2}{\circle*{.15}}
\matrixput(2,0)(1,0){2}(0,1){3}{\circle*{.15}}
\put(4,0){\circle*{.15}}

\dottedline{.2}(0,4)(2,4)(2,0)(4,0)(0,4)
\thicklines
\path(0,3)(2,0)
\dottedline{.12}(2,4)(4,1)

\multiput(0.06,3.06)(2,-3){2}{$\z_1$}
\put(1.06,2.06){$\z_2$}
\multiput(0.06,4.06)(2,-3){2}{$\z_3$}
\multiput(1.06,3.06)(2,-3){2}{$\z_4$}
\put(2.06,2.06){$\z_5$}
\multiput(1.06,4.06)(2,-3){2}{$\z_6$}
\multiput(2.06,3.06)(2,-3){2}{$\z_7$}
\put(3.06,2.06){$\z_8$}
\multiput(2.06,4.06)(2,-3){2}{$\z_9$}
\put(3.06,3.06){$\z_{10}$}
\end{picture}
\end{center}
\caption{\label{iv2eq} (2,-3)-periodic initial values}
\end{figure}
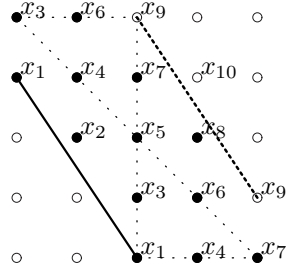

We specify initial values between two parallel lines that
squeeze the stencil as in Figure \ref{iv2eq}.

Clearly, given the 8 values $\z_1,\z_2,\ldots,\z_8$
at the black dots one is able to determine the value $\z_9$ at the white dots on the fat-dotted
line, using the HADT equation. Thus we get an eight dimensional mapping:
\be \label{mq2-3}
\begin{split}
\z_i &\mapsto \z_{i+1}, \quad 1\leq i <8, \\
\z_8 &\mapsto \frac{\z_6^2}{\z_4^2}\left(\frac{\z_1\z_7}{\z_3}-\z_5\right)+\frac{\z_7\z_5}{\z_3}.
\end{split}
\ee

To be able to evaluate monodromy matrix
\[
\cL=M_{1,0}^{-1}L_{0,1}M_{0,1}^{-1}M_{0,2}^{-1}L_{-1,3},
\]
we first have to determine the value $x_{10}$. Using the stencils of the Lax matrices, see Figures \ref{SL} and \ref{SM},
it seems we also need to to determine the value $x_0$, as both $L_{0,1}$ and $M_{0,1}^{-1}$ depend on $x_0$. However, their
 product does not depend on $x_0$. The characteristic polynomial is given by
\[
\det(\mu {\rm I} - \cL) = \mu^2((\mu-\lambda)^2 +(\mu-\lambda)I_2 - \mu JJ^\prime + 1) + (\mu-\lambda)^3 - \mu(\mu - \lambda)^2 I_1
\]
where
\begin{eqnarray*}
I_1 &=& \frac{\z_4\z_5}{\z_3\z_6}+\frac{\z_3\z_8}{\z_5\z_6}+\frac{\z_1\z_6}{\z_3\z_4}+\frac{\z_2\z_7}{\z_4\z_5}, \\
I_2 &=& \frac{\z_2\z_7}{\z_3\z_6}\left(\frac{\z_1\z_6^2}{\z_4^2\z_5}+\frac{\z_3^2\z_8}{\z_4\z_5^2}+1 \right)
+\frac{\z_1\z_8}{\z_4\z_5}-\frac{\z_6\z_2}{\z_4^2}-\frac{\z_3\z_7}{\z_5^2}, \\
J &=& \frac{\z_1\z_7-\z_5\z_3}{\z_4^2},
\end{eqnarray*}
where $J$ is a $2$-integral of the mapping. The four integrals $I_1,I_2,JJ^\prime$ and $J+J^\prime$ are functionally
independent.

\subsubsection{General $\s$-reduction}
By imposing $\s$-periodicity, that is,
\[
\Deltas_{l,m}=\Deltas_{l+s_1,m+s_2}.
\]
our P$\Delta$E reduces to a system of $r=\text{gcd}(s_1,s_2)$ O$\Delta$Es.
We perform so-called $\s$-reduction, which is basically a change of variables $(l,m)\rightarrow (n,p)$.

\begin{definition} We define $a,b,c,d\in\N$ and $\epsilon=\pm1$ by $a/b=|s_1/s_2|=\epsilon s_1/s_2$; gcd$(a,b)=1$;
$b=0 \Rightarrow c=0,d=1$; $b=1 \Rightarrow c=1,d=0$; $b>1 \Rightarrow c$ is the smallest positive integer such
that $bc-ad=1$. Now we specify
\be \label{np}
n_{l,m}:=b l-\epsilon a m, \qquad p_{l,m}:=-d l+\epsilon c m \text{ mod } r,
\ee
\end{definition}
The variable $n$ tells us on which line with direction $(a,\epsilon b)$ the point is and $p$
distinguishes the $r$ inequivalent points on each line, see also \cite{K}.
The important properties are
\[
(n,p)_{l+a,m+\epsilon b}=(n,p+1)_{l,m},\qquad (n,p)_{l+c,m+\epsilon d}=(n+1,p)_{l,m}
\]
where the latter shift is the standard way to update the initial values.

We adopt the notation of \cite{K}, where reduced variables carry one upper and one lower index, that is,
the reduction will be denoted $\Deltas_{l,m} \rightarrowtail \Deltas^p_n$.
When specifying (periodic) initial value problems in terms of reduced variables we specify sets of
variables $\Deltas_n^p$ where $n$ runs over a specific range and $p$ runs over the full range $p\in\N_r:=\{0,1,\ldots,r-1\}$,
cf. \cite{K}.

The range of $n$ is determined by the differences associated to each region. For the HADT equation,
when $\s\in R^1$ (where $\epsilon=-1$, see Figure \ref{scr}) the range of $n$ will be from $0$ to $n(d^1)-1 = b\cdot 0 + a\cdot 4 -1$,
and when $\s\in R^2$ we take $0\leq n < 4(b+\epsilon a) -1$, using the differences (\ref{eq:diffs}).

Thus we obtain
\begin{proposition} The HADT equation admits a well-posed $\s$-periodic initial
value problem if $(a,b)$ is not equal to $(1,0)$ or $(1,-2)$.
\begin{itemize}
\item For $\s\in R^1$ the set $\{\Deltas^p_n: n\in\N_{4a}, p\in\N_r\}$ provides a well-posed initial value problem
of dimension $4\abs{s_1}$.
\item For $\s\in R^2$ the set $\{\Deltas^p_n: n\in\N_{4(b+\epsilon a)}$, $p\in\N_r\}$ provides a well-posed initial
value problem of dimension $4\abs{s_1+s_2}$.
\end{itemize}
\end{proposition}

{\bf Remark:} In section \ref{s2-1s} the initial values $\{\Deltas^0_0,\ldots,\Deltas^0_7\}$ are conveniently
denoted $\Deltas^0_n=\z_{8-n}$ and the mapping is the backward shift $n\mapsto n-1$, or $(l,m)\mapsto(l-1,m)$.
In section \ref{s2-3s} the initial values $\{\Deltas^0_0,\ldots,\Deltas^0_7\}$ are denoted $\Deltas^0_n=\z_{n+1}$
and the mapping is the standard one $n\mapsto n+1$, that is $(l,m)\mapsto(l+1,m-1)$.

\subsection{Initial value problems for the system of lattice equations (\ref{eq:HS2})} \label{ss2}

The stencils of the system (\ref{eq:HS2}) are depicted in Figure \ref{SBH}.
The $S$-directions in the $S$-polygons of the equations are $(0,1)$, $(1,0)$, $(1,-1)$, and $(1,-2)$.
Therefore, we distinguish four regions in the plane, as in Figure \ref{DRBH}.\\[5mm]

\noindent
\parbox{60mm}{
\begin{center}
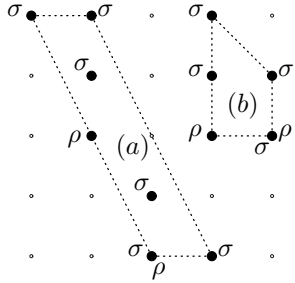

\setlength{\unitlength}{8mm}
\begin{picture}(4,4)(0,0)
\matrixput(0,0)(1,0){5}(0,1){5}{\circle{.05}}
\multiput(3,2)(0,1){3}{\circle*{.15}}
\multiput(4,2)(0,1){2}{\circle*{.15}}

\multiput(0,4)(1,-1){2}{\circle*{.15}}
\multiput(1,2)(1,-1){3}{\circle*{.15}}
\multiput(1,4)(1,-4){2}{\circle*{.15}}
\thinlines
\dottedline{.10}(0,4)(2,0)(3,0)(1,4)(0,4)
\dottedline{.10}(3,4)(3,2)(4,2)(4,3)(3,4)
\put(3.25,2.4){($b$)}
\put(1.4,1.7){($a$)}

\put(2.6,4){$\Deltas$}
\put(2.6,3){$\Deltas$}
\put(2.6,2){$\Thetar$}
\put(4.1,3){$\Deltas$}
\put(4.1,2){$\Thetar$}
\put(3.7,1.7){$\Deltas$}

\put(-.4,4){$\Deltas$}
\put(1.1,4){$\Deltas$}
\put(.7,3.1){$\Deltas$}
\put(.6,1.9){$\Thetar$}
\put(1.7,1.1){$\Deltas$}
\put(1.6,0){$\Deltas$}
\put(2,-.3){$\Thetar$}
\put(3.1,0){$\Deltas$}
\end{picture}
\captionof{figure}{\label{SBH} The 5-point and 7-point stencils of system (\ref{eq:HS2}), and their convex hulls}
\end{center}}
\hfill
\parbox{60mm}{
\setlength{\unitlength}{8mm}
\begin{center}
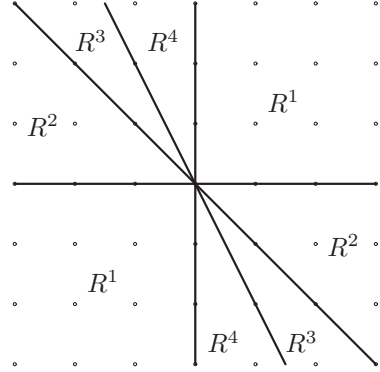

\begin{picture}(6,6)(0,0)
\matrixput(0,0)(1,0){7}(0,1){7}{\circle{.05}}
\thicklines
\path(0,3)(6,3)
\path(3,0)(3,6)
\path(0,6)(6,0)
\path(1.5,6)(4.5,0)
\put(4.2,4.2){$R^1$}
\put(5.2,1.8){$R^2$}
\put(4.5,.2){$R^3$}
\put(3.2,.2){$R^4$}
\put(1.2,1.2){$R^1$}
\put(.2,3.8){$R^2$}
\put(1,5.2){$R^3$}
\put(2.2,5.2){$R^4$}
\end{picture}
\captionof{figure}{\label{DRBH} Distinguished regions for system (\ref{eq:HS2})}
\end{center}}

For each of the distinguished regions $R^i$ we will project each stencil of the system onto
a line with direction $\s\in R^i$. On these lines we will indicate (schematically) the ranges
of $n$-values for both field variables $\Deltas, \Thetar$ (our ranges start at 0). This procedure
is similar to the scalar case, where the difference $d^i$ is equivalent to a range for $n$.
For every equation in a system we need a range (and hence a difference) for every field, and also
their relative positions. Moreover, once we have this data, there is still some freedom
left how to position the equations with respect to each other. In fact, as we will show in Appendix
\ref{appendix4}, we have to split region $R^4$ into two regions where the two equations are positioned
differently with respect to each other. Also we will see that for regions $R^2$ and $R^3$ we may take
the same initial values. However, updating the initial values is done differently in those regions.

\begin{figure}[h]
\setlength{\unitlength}{8mm}
\begin{center}
\begin{picture}(8,2.5)(0,0)
\path(0,0)(8,0)
\multiput(0,0)(2,0){5}{\path(0,-.1)(0,.1)}
\thicklines
\path(0,1)(8,1)
\path(0,2)(4,2)
\multiput(0,1)(0,1){2}{\circle*{.15}}
\multiput(4,1)(0,1){2}{\circle*{.15}}
\put(2,2){\circle*{.15}}
\put(6,1){\circle*{.15}}
\put(8,1){\circle*{.15}}
\put(-.1,.2){$0$}
\put(1.8,.2){$2a$}
\put(3.4,.2){$2a+b$}
\put(5.4,.2){$4a+2b$}
\put(7.4,.2){$4a+3b$}
\put(-.3,1.1){$\Deltas$}
\put(-.3,2.1){$\Deltas$}
\put(1.7,2.1){$\Thetar$}
\put(3.7,1.1){$\Thetar$}
\put(3.6,2.1){$\Deltas\ \Thetar$}
\put(5.7,1.1){$\Thetar$}
\put(7.7,1.1){$\Deltas$}
\end{picture}
\captionof{figure}{\label{SRBH} $\s$-Reduction for system (\ref{eq:HS2}), with $\s\in\N^2$}
\end{center}
\end{figure}
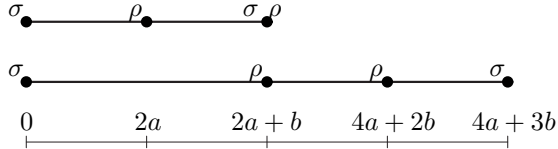

For $\s \in R^1$ the ranges are given in Figure \ref{SRBH}.

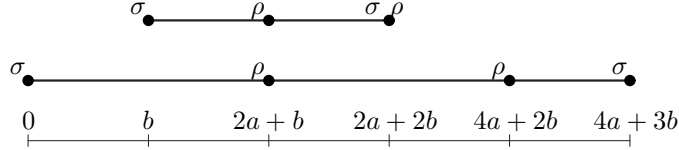
\begin{figure}[h]
\setlength{\unitlength}{8mm}
\begin{center}
\begin{picture}(10,2.5)(0,0)
\path(0,0)(10,0)
\multiput(0,0)(2,0){6}{\path(0,-.1)(0,.1)}
\thicklines
\path(0,1)(10,1)
\path(2,2)(6,2)
\put(0,1){\circle*{.15}}
\put(2,2){\circle*{.15}}
\multiput(4,1)(0,1){2}{\circle*{.15}}
\put(6,2){\circle*{.15}}
\put(8,1){\circle*{.15}}
\put(10,1){\circle*{.15}}
\put(-.1,.2){$0$}
\put(1.9,.2){$b$}
\put(3.4,.2){$2a+b$}
\put(5.4,.2){$2a+2b$}
\put(7.4,.2){$4a+2b$}
\put(9.4,.2){$4a+3b$}
\put(-.3,1.1){$\Deltas$}
\put(1.7,2.1){$\Deltas$}
\put(3.7,2.1){$\Thetar$}
\put(3.7,1.1){$\Thetar$}
\put(5.6,2.1){$\Deltas\ \Thetar$}
\put(7.7,1.1){$\Thetar$}
\put(9.7,1.1){$\Deltas$}
\end{picture}
\captionof{figure}{\label{SRBH2} Alternative $\s$-Reduction for system (\ref{eq:HS2}), with $\s\in\N^2$}
\end{center}
\end{figure}

We now have to find a range of $n$ for each variable $\Deltas,\Thetar$ such that each equation
can be used to update one of the variables. Moreover, we want to minimise the number of field values
that have to be calculated to update both fields. Therefore, we move the first equation in Figure \ref{SRBH}
over a distance $b$ with respect to the other, see Figure \ref{SRBH2}.

For system (\ref{eq:HS2}) with $\s\in R^1$ one can see that if we take the range $0\leq n < 4a+3b$
for $\Deltas$ and the range $2a+b\leq n
< 2a+2b$ for $\Thetar$, then using the equation (\ref{eq:HS2a}) we can determine the values of
$\Thetar$ at $2a+2b \leq n \leq 4a+2b$. Subsequently, the equation (\ref{eq:HS2b}) can be used
to calculate the values of $\Deltas$ at $n=4a+3b$. We have obtained the first item in the
following proposition, whose complete proof is given in Appendix \ref{appendix4}.

\begin{proposition} \label{PBH}
Let $\s\in\Z^2$ be such that $(a,b)$ is not equal to $(1,0)$ or $(1,-2)$, and let $R^i$ be as in Figure
\ref{DRBH}.
Then system (\ref{eq:HS2}) admits a well-posed $\s$-periodic initial value problem.
\begin{itemize}
\item
For $\s\in R^1$ the set
\[
\{\Deltas^p_n,\Thetar^p_m: n\in\N_{4a+3b}, m-2a-b\in\N_{b}, p\in\N_r\}
\]
provides a well-posed initial value problem of dimension $4\abs{s_1+s_2}$.
\item
For $\s\in R^2 \cup R^3$ (including the boundary between $R^2$ and $R^3$) the set
\[
\{\Deltas^p_n, \Thetar^p_m: n\in\N_{4a-b}, m\in\N_{b}, p\in\N_r\}
\]
provides a well-posed initial value problem of dimension $4\abs{s_1}$.
\item
For $\s\in R^4$, $b\leq 3a$, the set
\[
\{\Deltas^p_n, \Thetar^p_m: n\in\N_{3b-4a}, m+5a-2b\in\N_{b}, p\in\N_r\}
\]
provides a well-posed initial value problem of dimension $4\abs{s_1+s_2}$.
\item
For $\s\in R^4$, $b\geq 3a$, the set
\[
\{\Deltas^p_n, \Thetar^p_m: n\in\N_{3b-4a}, m+4a-b\in\N_{b}, p\in\N_r\}
\]
provides a well-posed initial value problem of dimension $4\abs{s_1+s_2}$.
\end{itemize}
\end{proposition}

\def\y{{y}}
\subsubsection{(2,-1)-reduction of system (\ref{eq:HS2})}

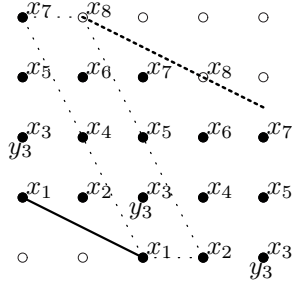
\begin{figure}[h]
\setlength{\unitlength}{8mm}
\begin{center}
\begin{picture}(4,4)(0,0)
\matrixput(0,0)(1,0){5}(0,1){5}{\circle{.15}}
\multiput(0,3)(1,0){3}{\circle*{.15}}
\multiput(2,0)(1,0){3}{\circle*{.15}}
\put(0,4){\circle*{.15}}
\matrixput(0,1)(1,0){5}(0,1){2}{\circle*{.15}}

\dottedline{.2}(0,4)(2,0)(3,0)(1,4)(0,4)
\thicklines
\path(0,1)(2,0)
\dottedline{.12}(1,4)(4,2.5)

\multiput(0.06,1.06)(2,-1){2}{$\z_1$}
\multiput(1.06,1.06)(2,-1){2}{$\z_2$}
\multiput(0.06,2.06)(2,-1){3}{$\z_3$}
\multiput(1.06,2.06)(2,-1){2}{$\z_4$}
\multiput(0.06,3.06)(2,-1){3}{$\z_5$}
\multiput(1.06,3.06)(2,-1){2}{$\z_6$}
\multiput(0.06,4.06)(2,-1){3}{$\z_7$}
\multiput(1.06,4.06)(2,-1){2}{$\z_8$}
\multiput(-.24,1.73)(2,-1){3}{$\y_3$}
\end{picture}
\end{center}
\caption{\label{iv1sy} (2,-1)-periodic initial values}
\end{figure}

We specify initial values between two parallel lines that
squeeze the $\Deltas$-stencil, as in Figure \ref{iv1sy}. The initial values are denoted $\Deltas^0_n=\z_{n+1}$ and
$\Thetar^0_n=\y_{n+1}$.
Clearly, from these initial values we can calculate the values $\y_4$ and $\y_2$, $\y_1$, using equation (\ref{eq:HS2b}).
Then we know enough $\Thetar$-values to be able to calculate, using equation (\ref{eq:HS2a}), $\z_8$, which (to our surprise) does
not depend on $\y_3$. Therefore, the $\Deltas$-field is determined by a 7-dimensional mapping, namely
\be \label{mb2-1}
\begin{split}
\z_i &\mapsto \z_{i+1}, \quad 1\leq i <7, \\
\z_7 &\mapsto \frac{\z_2\z_7}{\z_1}-\frac{\z_2\z_5\z_6}{\z_1\z_4}-\frac{\z_3\z_4\z_7}{\z_1\z_5}
-\frac{\z_3^2\z_6^2}{\z_1\z_4\z_5}.
\end{split}
\ee

We know enough values to determine the monodromy matrix $\cL=L_0^1(M_0^1)^{-1}L_0^1$.
Its characteristic polynomial is given by
\[
\det(\mu {\rm I} - \cL) = \mu^2((\mu-\lambda)^2 + (\mu-\lambda)I_1+1) - \lambda( \mu^2 JJ^\prime +\mu I_1 +1),
\]
where
\begin{eqnarray*}
I_1&=&-\frac{\z_3\z_4}{\z_1\z_6}+\frac{\z_3\z_6}{\z_4\z_5}
-\frac{\z_4\z_5}{\z_7\z_2}+\frac{\z_1\z_4}{\z_2\z_3}+\frac{\z_5\z_2}{\z_3\z_4}
+\frac{\z_4\z_7}{\z_5\z_6}+\frac{\z_5\z_2}{\z_1\z_6}\nn \\
&&+\frac{\z_6\z_3}{\z_7\z_2} -\frac{\z_5^2\z_2}{\z_7\z_1\z_4}-\frac{\z_6\z_3^2}{\z_7\z_1\z_4}\\
J&=&\frac{\z_1\z_7-\z_5\z_3}{\z_2\z_6}+\frac{\z_3^2}{\z_2\z_4}+\frac{\z_5^2}{\z_4\z_6}.
\end{eqnarray*}
The integrals $I_1,JJ^\prime,J+J^\prime$ are functionally independent.

\subsubsection{(2,-3)-reduction of system (\ref{eq:HS2})}
\begin{figure}[h]
\setlength{\unitlength}{8mm}
\begin{center}
\begin{picture}(4,4)(0,0)
\matrixput(0,0)(1,0){5}(0,1){5}{\circle{.15}}
\multiput(0,4)(1,-1){3}{\circle*{.15}}
\multiput(0,3)(1,-1){4}{\circle*{.15}}
\put(2,0){\circle*{.15}}

\dottedline{.2}(0,4)(2,0)(3,0)(1,4)(0,4)
\thicklines
\path(0,3)(2,0)
\dottedline{.12}(1,4)(3.66,0)
\multiput(0.06,3.06)(2,-3){2}{$\z_1$}
\multiput(0.06,4.06)(2,-3){2}{$\z_3$}
\multiput(1.06,3.06)(2,-3){2}{$\z_4$}
\put(1.06,2.06){$\z_2$}
\put(2.06,2.06){$\z_5$}
\multiput(-.24,2.73)(2,-3){2}{$\y_1$}
\multiput(-.24,3.73)(2,-3){2}{$\y_3$}
\put(.76,1.73){$\y_2$}
\end{picture}
\end{center}
\caption{\label{iv2sy} (2,-3)-periodic initial values}
\end{figure}
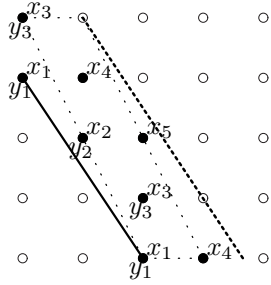

We specify initial values between two parallel lines that
squeeze the $\Deltas$-stencil, as in Figure \ref{iv2sy}.
From these initial values we can calculate the $\Deltas$-value ($\z_6$) on the dashed line.
And once we have done that we can calculate $\y_4$. We find an
8-dimensional mapping, namely
\be \label{mb2-3}
\begin{split}
\z_i &\mapsto \z_{i+1}, \quad 1\leq i <5, \\
\z_5 &\mapsto \z_5^\prime = \frac{\z_4\z_3+\y_1\z_4-\z_3\y_2}{\z_1},\\
\y_i &\mapsto \y_{i+1}, \quad i=1,2, \\
\y_3 &\mapsto \frac{\z_4\z_5+\y_1\z_5^\prime}{\z_3}.
\end{split}
\ee

One has to calculate quite a few other points to be able to evaluate the monodromy matrix,
whose characteristic polynomial is given by\footnote{The factorization of $I_2+I_1+1$ where $I_2$ is
the coefficient of $\mu^2\lambda$ is due to Claude Viallet \cite{CV}.}
\[
\det(\mu {\rm I} - \cL) = \mu^2( (\mu-\lambda)^2+\mu I_1 + \lambda (JJ^\prime-I_1-1) + 1) - \lambda^3,
\]
where 
\begin{eqnarray*}
I_1&=&-\frac{x_3x_2}{x_1x_4}+\frac{y_3y_1}{x_5x_3}+\frac{y_3y_2}{x_5x_4}+\frac{y_2y_1}{x_3x_4}-\frac{y_2^2}{x_4^2}-\frac{x_5x_1}{x_3^2}+\frac{x_3x_2y_2}{x_1x_4^2}\\
&&-\frac{x_3x_4}{x_5x_2}-\frac{x_2y_1}{x_1x_4}-\frac{x_3y_2}{x_5x_2}+\frac{x_3x_4y_3}{x_5^2x_2}-\frac{y_1^2}{x_3^2}-\frac{y_3^2}{x_5^2}\\
J&=&\frac{x_1x_5y_2}{x_4^2x_2}-\frac{y_3x_1}{x_4x_2}+\frac{x_5x_1}{x_4x_2}-\frac{x_5x_3}{x_4^2}.
\end{eqnarray*}
The integrals $I_1, JJ^\prime, J+J^\prime$ are functionally independent.


\subsection{Initial value problems for the QQD scheme} \label{ss3}
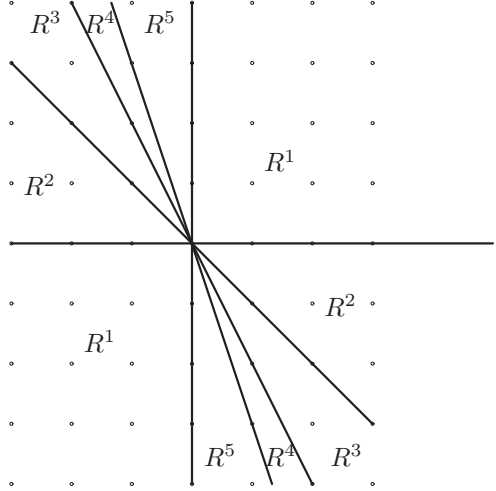
\begin{figure}[h]
\setlength{\unitlength}{8mm}
\begin{center}
\begin{picture}(6,8)(0,0)
\matrixput(0,0)(1,0){7}(0,1){9}{\circle{.05}}
\thicklines
\path(0,4)(8,4)
\path(3,0)(3,8)
\path(0,7)(6,1)
\path(1,8)(5,0)
\path(1.66,8)(4.33,0)
\put(4.2,5.2){$R^1$}
\put(5.2,2.8){$R^2$}
\put(5.3,.3){$R^3$}
\put(4.2,.3){$R^4$}
\put(3.2,.3){$R^5$}
\put(1.2,2.2){$R^1$}
\put(.2,4.8){$R^2$}
\put(.3,7.5){$R^3$}
\put(1.2,7.5){$R^4$}
\put(2.2,7.5){$R^5$}
\end{picture}
\caption{\label{FRQQD} Distinguished regions for the QQD scheme}
\end{center}
\end{figure}

We define regions in the plane as follows
\begin{eqnarray*}
R^1&=&\{\s\in\Z^2:a\geq 0,b>0,\epsilon=1\}, \\
R^2&=&\{\s\in\Z^2:0<b\leq a,\epsilon=-1\}, \\
R^3&=&\{\s\in\Z^2:a\leq b<2a,\epsilon=-1\}, \\
R^4&=&\{\s\in\Z^2:2a<b\leq 3a,\epsilon=-1\}, \\
R^5&=&\{\s\in\Z^2:3a\leq b,\epsilon=-1\},
\end{eqnarray*}
see Figure \ref{FRQQD}.
We prove Proposition \ref{PQQD} in Appendix \ref{appendix5}.
\begin{proposition} \label{PQQD}
The QQD scheme (\ref{eq:QQD}) admits a well-posed $\s$-periodic initial value problem if
$(a,b)$ is not equal to $(1,0)$ or $(1,-2)$.
\begin{itemize}
\item
For $\s\in R^1$ the set
\[
\{\A^p_n,\B^p_m,\C^p_l: n\in\N_{3a+2b}, m-2a\in\N_{a+b}, l-2a-b\in\N_{b}, p\in\N_r\}
\]
provides a well-posed initial value problem of dimension $4\abs{s_1+s_2}$.
\item
For $\s\in R^2$ the set
\[
\{\A^p_n,\B^p_m,\C^p_l: n-b \in \N_{3a-b}, m\in\N_{a}, l \in\N_{b}, p\in\N_r\}
\]
provides a well-posed initial value problem of dimension $4\abs{s_1}$.
\item
For $\s\in R^3$ the set
\[
\{\A^p_n,\B^p_m,\C^p_l: n+a-b \in \N_{3a-b}, m\in\N_{a}, l\in\N_{b}, p\in\N_r\}
\]
provides a well-posed initial value problem of dimension $4\abs{s_1}$.
\item
For $\s\in R^4$ the set
\[
\{\A^p_n,\B^p_m,\C^p_l: n-3a+b \in \N_{2b-3a}, m\in\N_{b-a}, l\in\N_{b}, p\in\N_r\}
\]
provides a well-posed initial value problem of dimension $4\abs{s_1+s_2}$.
\item
For $\s\in R^5$ the set
\[
\{\A^p_n,\B^p_m,\C^p_l: n-2a \in \N_{2b-3a}, m-a\in\N_{b-a}, l \in\N_{b}, p\in\N_r\}
\]
provides a well-posed initial value problem of dimension $4\abs{s_1+s_2}$.
\end{itemize}
\end{proposition}

In the following reductions we adopt the notation $\A^0_n=x_n,\B^0_n=y_n,\C^0_n=z_n$.
\subsubsection{(2,-1)-reduction of the QQD scheme}
Updating the initial values as described in Appendix \ref{appendix5} yields the following
mapping
\be
\begin{split}
x_i &\mapsto x_{i+1}, \quad 1\leq i <5, \\
x_5 &\mapsto \frac{x_1y_1}{y_1^\prime}, \\
y_0 &\mapsto y_1, \\
y_1 &\mapsto y_1^\prime = x_1+y_0-x_5-z_0+\frac{x_5x_4z_0}{x_1x_2}, \\
z_0 &\mapsto \frac{x_4z_0}{x_1}.
\end{split}
\ee

The characteristic polynomial of the monodromy matrix is given by
\[
\det(\mu {\rm I} - \cL) = \mu^2((\mu-\lambda)^2 + (\mu-\lambda)I_1 -\lambda JJ^\prime + I_2 ) + \lambda(\mu I_3 - I_4),
\]
where
\[
\begin{split}
I_1 &= (y_0-z_0+x_1)\frac{z_0x_4}{x_1}+y_1(z_0+x_5-y_0), \\
I_2 &= z_0x_4x_5y_1, \\
I_3 &= (z_0x_2-x_5y_1-x_2x_1-y_0x_2)x_3x_4-x_1y_1(x_5x_4+x_2x_5+x_2x_3), \\
I_4 &= x_1x_2x_3x_4x_5y_1, \\
J &= x_1+x_3+y_0-z_0.
\end{split}
\]
The integrals $I_1,I_2,I_3,I_4,I_5=JJ^\prime,I_6=J+J^\prime$ are functionally independent.

\subsubsection{(2,-3)-reduction of the QQD scheme}
Updating the initial values as described in Appendix \ref{appendix5} yields the following
mapping
\be
\begin{split}
x_i &\mapsto x_{i+1}, \quad i=1,2, \\
x_3 &\mapsto x_3^\prime=\frac{y_1x_1}{y_0}, \\
y_0 &\mapsto y_1 \\
y_1 &\mapsto y_0+z_2-z_0, \\
z_i &\mapsto z_{i+1}, \quad i=0,1, \\
z_2 &\mapsto \frac{x_3^\prime z_0}{x_3}.
\end{split}
\ee

The characteristic polynomial of the monodromy matrix is given by
\[
\begin{split}
\det(\mu {\rm I} - \cL) = &\mu^2((\mu-\lambda)^2 +  \mu((I_4-I_2+1)I_1-I_3-JJ^\prime) + \lambda(I_3+(2I_2-3)I_1) )\\
&+ \lambda^2(\mu(3-I_2)-\lambda)I_1,
\end{split}
\]
where
\begin{eqnarray*}
I_1 &=& \frac{x_2x_1x_3}{y_0}, \\
I_2 &=& \frac{z_2}{x_3}+\frac{y_0}{x_3}+\frac{y_0}{x_1}+\frac{y_1}{x_2}, \\
I_3 &=& x_2z_0-x_2y_0-x_2z_2+z_1x_3-x_3y_1-y_1x_1, \\
I_4 &=& \frac{z_0z_1z_2}{x_3}, \\
J &=& y_0-z_0
\end{eqnarray*}

\section{Dimensional reduction} \label{5}
\setcounter{equation}{0}
In this section we present the explicit formulae for the reduced
variables, required to implement the dimensional reduction of the mappings
obtained in section 4 from periodic reduction from the HADT equation
and from the system (\ref{eq:HS2}). These formulae arise from
the symmetries of the partial difference equations.  In certain cases
an extra scaling symmetry of the integrals has to be used to obtain
the dimensional reduction.

Whereas the dimensions of the $\s$-reductions from the HADT equation, of  the intermediate system, and of the QQD scheme all coincide, these reductions differ both with regard to the number of integrals we found in the previous section and with regard to the number of symmetries we will encounter in the present  section.
The question whether the $\s$-reductions of the various equations are related, e.g. through a change of variables, is left open.

\def\w{{w}}
\subsection{The HADT equation}
Equation (\ref{eq:quadd}) is invariant under the following continuous symmetries:
$u\mapsto u\epsilon$ and $u\mapsto u\epsilon^{(-1)^l}$. The first yields a
symmetry of the mappings, whereas the second gives us a $2$-symmetry of the
mappings. Their generators are
\[
\sum_{i=1}^8 \z_i \frac{\partial}{\partial \z_i}, \quad
\sum_{i=1}^8 (-1)^i\z_i \frac{\partial}{\partial \z_i}.
\]
A set of joint invariants of these vector fields is given by
\[
\{\w_1=\frac{\z_1\z_4}{\z_2\z_3}, \w_2=\frac{\z_2\z_5}{\z_3\z_4}, \w_3=\frac{\z_3\z_6}{\z_4\z_5},
\w_4=\frac{\z_4\z_7}{\z_5\z_6}, \w_5=\frac{\z_5\z_8}{\z_6\z_7}, \w_6=\frac{\z_6\z_1}{\z_7\z_8}\}.
\]
In these variables the $\s=(2,-1)$ mapping (\ref{mq2-1}) reads
\be
\begin{split}
\w_i &\mapsto \w_{i+1}, \quad 1\leq i <5, \\
\w_5 &\mapsto \frac{f}{\w_2\w_3\w_4\w_5}, \quad f=\w_2-\w_5+(\w_3-\w_2)\w_4+\w_1\w_2\w_3\w_4\w_5, \\
\w_6 &\mapsto \frac{\w_2\w_4\w_6}{\w_1 f}.
\end{split}
\ee
which has three functionally independent invariants
\bse\begin{align}
I_1 &= \w_1+\w_5+\w_2\w_4(1-\w_1\w_3\w_5) + \frac{\w_3-1}{\w_2\w_3\w_4}, \\
I_2 &= \frac{(\w_1 \w_2 \w_3 \w_4 + \w_2 + \w_3- 1 )(\w_2 \w_3 \w_4 \w_5 + \w_3 + \w_4- 1)}{\w_2 \w_3 \w_4}, \\
I_3 &= \w_2+\w_3+\w_4-\w_4\w_2+\w_1\w_2\w_3\w_4\w_5.
\end{align}\ese
The 2-integral $J$ is scaled by the generator of the second symmetry and so it does not survive the reduction.
The $\s=(2,-3)$ mapping (\ref{mq2-3}) reduces to the six dimensional mapping
\be
\begin{split}
\w_i &\mapsto \w_{i+1}, \quad 1\leq i <5, \\
\w_5 &\mapsto \frac{g}{\w_4\w_5}, \quad g=\w_1\w_2\w_3+\frac{1}{\w_3} - \frac{1}{\w_4} \\
\w_6 &\mapsto \frac{\w_4\w_6}{\w_1\w_3 g}
\end{split}
\ee
and has three functionally independent invariants
\begin{align*}
I_1 &= \w_3(\w_1\w_2+\w_2\w_4+\w_4\w_5)+\frac{1}{\w_3}, \\
I_2 &= \w_2\w_3^2\w_4(\w_1\w_2+\w_1\w_5+\w_4\w_5)-\w_2\w_3-\w_3\w_4+\w_2\w_4, \\
JJ^\prime &= \w_3(\w_1\w_2\w_3\w_4-1)(\w_2\w_3\w_4\w_5-1).
\end{align*}

\subsection{The lattice system (\ref{eq:HS2})}
For the case $\s=(2,-1)$ the mapping has a symmetry and a 2-symmetry, which arise
from scaling symmetries of the system. Their generators are given by
\[
\sum_{i=1}^7 \z_i \frac{\partial}{\partial \z_i}, \quad \sum_{i=1}^7 (-1)^i \z_i \frac{\partial}{\partial \z_i}.
\]
The integrals admit an extra scaling, which is neither a symmetry, nor a $k$-symmetry, of the mapping, namely
\[
\z_1 \frac{\partial}{\partial \z_1} + \z_2 \frac{\partial}{\partial \z_2} - \z_5 \frac{\partial}{\partial \z_5}
-\z_6 \frac{\partial}{\partial \z_6} - 2 \z_7 \frac{\partial}{\partial \z_7}.
\]
Still, we can use joint invariants of these three vector fields,
\[
\w_1=\frac{\z_1\z_4}{\z_2\z_3}, \w_2=\frac{\z_5\z_2}{\z_3\z_4}, \w_3=\frac{\z_3\z_6}{\z_5\z_4}, \w_4=\frac{\z_4\z_7}{\z_5\z_6},
\]
to reduce the dimension of the mapping by 3 dimensions. This is due to the fact that
if we reduce the third vector field by the first two the resulting vector field
is a symmetry of the reduced mapping. The mapping (\ref{mb2-1}) reduces to a four dimensional mapping
\be
\begin{split}
\w_i &\mapsto \w_{i+1}, \quad 1\leq i <4, \\
\w_4 &\mapsto \frac{\w_2\w_4-\w_2-\w_3-\w_4}{\w_1\w_2\w_3\w_4},
\end{split}
\ee
which admits two functionally independent integrals
\begin{eqnarray*}
I_1&=&\w_1+\w_2+\w_3+\w_4+\frac{\w_1\w_3+\w_2\w_4-\w_1-\w_2-\w_3-\w_4}{\w_1\w_2\w_3\w_4}, \\
JJ^\prime&=&(\w_1+\w_2+\w_3+\w_4-\w_1\w_3-\w_1\w_4-\w_2\w_4)\frac{\w_1\w_2\w_3\w_4+\w_2+\w_3-1}{\w_1\w_2\w_3\w_4}.
\end{eqnarray*}

Similarly we find three reductions for the $\s=(2,-3)$ case. Here choosing variables
\[
\w_1=\frac{\y_1}{\z_3}, \w_2=\frac{\y_2}{\z_4}, \w_3=\frac{\y_3}{\z_5}, \w_4=\frac{\z_2\z_3}{\z_1\z_4}, \w_5=\frac{\z_3\z_4}{\z_2\z_5}
\]
reduces the mapping (\ref{mb2-3}) to the five dimensional mapping
\be
\begin{split}
\w_i &\mapsto \w_{i+1}, \quad i=1,2,4, \\
\w_3 &\mapsto \w_1+ \w_5^\prime, \\
\w_5 &\mapsto \w_5^\prime = \frac{1}{(1+\w_1-\w_2)\w_4\w_5}.
\end{split}
\ee
This mapping admits the two integrals
\begin{eqnarray*}
I_1&=&\w_1(\w_2-\w_1)+\w_2(\w_3-\w_2)+\w_3(\w_1-\w_3) +\w_4(\w_2-\w_1-1)\\
&& + \w_5(\w_3-\w_2-1)-\frac{1}{\w_4\w_5}, \\
JJ^\prime&=&(1+\w_2-\w_3-\w_4)\Big(1-\w_2+\w_3-\w_5-(\w_1-\w_3+\w_5)(\w_1-\w_2)\\
&&-\frac{1}{\w_4\w_5}\Big).
\end{eqnarray*}

We note that in both cases the function $J$ is no longer a 2-integral of the reduced mapping, and
that $J+J^\prime$ can not be expression in the reduced variables. Also, in the $\s=(2,-3)$ case
we have only two integrals for a five dimensional mapping which is not enough for complete integrability.
The reduction was done using 3 (scaling) symmetries of the integrals, but the function $J$ is only invariant
under one of them. It turns out it is also invariant under a linear combination of the
other two. Using this, we can then perform 2-reduction for mapping (\ref{mb2-3}), retaining all three
integrals. Choosing variables
\[
z_1 = \frac{x_2^2}{x_3 x_1},
z_2 = \frac{x_3^2}{x_4 x_2},
z_3 = \frac{x_4^2}{x_5 x_3},
z_4 = \frac{y_2}{x_4},
z_5 = \frac{y_3}{x_5},
z_6 = \frac{y_1 x_5}{y_2^2},
\]
the mapping reads
\be
\begin{split}
z_i &\mapsto z_{i+1}, \quad i=1,2,4 \\
z_3 &\mapsto z_3^\prime = z_1 z_2^2 z_3^2 (1+z_6 z_4^2 z_3-z_4) \\
z_5 &\mapsto z_3 (z_3^\prime + z_4^2 z_6)\\
z_6 &\mapsto \frac{z_4}{z_3^\prime z_5^2},
\end{split}
\ee
which has three invariants:
\[
z_6 z_3 z_4^2 (z_6 z_4^2 z_3+ z_1 z_2 - z_4 - z_5) + z_4^2 + z_5^2 - z_4 z_5 +
\frac{1}{z_1 z_2^2z_3} - (z_4 -1) z_1 z_2 - (z_5 -z_4 -1) z_3 z_2,
\]
and $J + J^\prime$ and $J J^\prime$, where $J = (1 + z_4 - z_5 - z_1 z_2)(z_1 z_2 z_3)^{-1}$.




\section{Acknowledgements}

Paul Spicer is funded by the research grant OT/08/033 of the research Council of Katholieke Universiteit Leuven.  Peter van der Kamp is funded by the Australian Research Council through the Centre of Excellence for Mathematics and Statistics of Complex Systems. Many thanks to Claude Viallet for providing the 2-integral for the mapping (\ref{mb2-3}).

\newpage

\appendix

\section{Some determinantal identities}\label{appendix1}
\setcounter{equation}{0}
\def\theequation{A.\arabic{equation}}

In the establishment of the recursive structure we need a number of determinantal identities, which we derive using the Sylvester Identity.
So we present a proof of the Sylvester identity, which was first presented by Kowalewski \cite{Kowalewski},
Bareiss \cite{Bareiss} and Malaschonok \cite{Mala1,Mala2}
and these seven proofs are presented together in \cite{Akritas}.

We consider an $(n+m)\times (n+m)$ matrix $R$ with elements $r_{ij}$ and determinant $|R|$, also written $\det (R)$.
Then we partition $R$ and factor by block triangularization such that
\be\label{eq:ACrel}
R=\left(\begin{array}{cc} A & B\\ C & D\end{array}\right)
=\left(\begin{array}{cc} A & 0\\ C &
\boldsymbol{1}\end{array}\right). \left(\begin{array}{cc}
\boldsymbol{1} & A^{-1}B\\ 0 &
D-CA^{-1}B\end{array}\right)
\ee
where $A$ is a nonsingular square matrix of order $n$, then
\be\label{eq:Rbloc}
|R|=|A|.| D-C A^{-1}B| .
\ee
If we multiply both sides by $|A|^{m-1}$, this becomes
\be
|A|^{m-1}|R|  =  ||A|(D-C A^{-1}B)|\nn
\ee
because the determinant on the right side of (\ref{eq:Rbloc}) is of order $m$.
We can reduce this equation further to
\be
|A|^{m-1}|R|= ||A| D-C \wt{A}B|,
\ee
since $A^{-1}=\frac{\wt{A}}{|A|}$ (where $\wt{A}$ represents the adjugate matrix of the inverse matrix $A^{-1}$),
and the determinant of $A$ is assumed to be $\neq 0$.
Specifying some entries in  (\ref{eq:ACrel}), taking $A$ to be
an $n\times n$ block and $D$ to be an $m\times m$ block, we have
the formula:
\begin{equation}\label{eq:genSylv}
(\det(A))^{m-1} \left| \begin{array}{ccccc}
       A      & |  &  b_{1}   &  \ldots  &  b_{m}  \\
         -     & +  &  -          &  -           &  -         \\
c^{t}_{1}  &  |  &              &             &             \\
\vdots     & |   &              &    D       &             \\
c^{t}_{m} & |   &             &              &
\end{array}\right|
={\det}_{m\times m}\left\{ {\det}_{n\times
n}(A)D_{ij}-(c^{t}_{i}\,\wt{A}\,b_{j})\right\}_{i,j=1,\cdots,m}
\end{equation}
in which the full matrix is supplemented with $m$ $n$-component
column vectors $b_{i}$ and $m$ $n$-component row-vectors
$c^{t}_{i}$.
If we consider the case $m=2$ ie. the removal of two rows and columns, then we get then determinant identity
{\small\begin{eqnarray*}
(\det(A))\left| \begin{array}{cccc}
          A   &  |   &    b_{1}  &  b_{2}  \\
          -    &  +  &  -           &  -   \\
c^{t}_{1}  &  |  &  d_{11}  &  d_{12}  \\
c^{t}_{2}  &  |  &  d_{21}  &  d_{22}
\end{array}\right|
 & = & {\det}_{2\times 2}\left\{ (\det(A))\left( \begin{array}{cc}
d_{11}  &  d_{12}  \\
d_{21}  &  d_{22}
\end{array}\right)
-\left( \begin{array}{cc}
c^{t}_{1}\wt{A}b_{1}  &  c^{t}_{1}\wt{A}b_{2}  \\
c^{t}_{2}\wt{A}b_{1}  &  c^{t}_{2}\wt{A}b_{2}
\end{array}\right)\right\}  \\
& = & [\det(A)d_{11}-c^{t}_{1}\wt{A}b_{1}]\,  [\det(A)d_{22}-c^{t}_{2}\wt{A}b_{2}] \\
&  &-[\det(A)d_{21}-c^{t}_{2}\wt{A}b_{1}]\, [\det(A)d_{12}-c^{t}_{1}\wt{A}b_{2}],
\end{eqnarray*}}
which can be symbolically written as:
\begin{eqnarray}
&& \left|
\begin{picture}(59,32)
\end{picture}
\right| \left|
\begin{color}{red}
\begin{picture}(59,32)
\put(48,-26){\rule{.3mm}{2cm}}\put(53,-26){\rule{.3mm}{2cm}}
\put(1,-21){\rule{2cm}{.3mm}}\put(1,-16){\rule{2cm}{.3mm}}
\end{picture}
\end{color}
\right|
= \left|
\begin{color}{red}
\begin{picture}(59,32)
\put(48,-26){\rule{.3mm}{2cm}}\put(1,-16){\rule{2cm}{.3mm}}
\end{picture}
\end{color}
\right|\times\left|
\begin{color}{red}
\begin{picture}(59,32)
\put(53,-26){\rule{.3mm}{2cm}}\put(1,-21){\rule{2cm}{.3mm}}
\end{picture}
\end{color}
\right| \nn  \\
&-&\left|
\begin{color}{red}
\begin{picture}(59,32)
\put(48,-26){\rule{.3mm}{2cm}}\put(1,-21){\rule{2cm}{.3mm}}
\end{picture}
\end{color}
\right|\times\left|
\begin{color}{red}
\begin{picture}(59,32)
\put(53,-26){\rule{.3mm}{2cm}}\put(1,-16){\rule{2cm}{.3mm}}
\end{picture}
\end{color}
\right|\quad\label{eq:othSylv}
\end{eqnarray}
(where the red lines denote rows and columns omitted from the original determinant).
It is then necessary to reorder the position of the row and column to tailor the identity to our requirements.

While (\ref{eq:othSylv}) is the key identity by which the recurrence
structure for ordinary one-variable orthogonal polynomials is
obtained, for the elliptic two-variable orthogonal polynomials we need (in addition to
(\ref{eq:othSylv})), determinantal identities involving the
simultaneous removal of more than two rows and columns. Thus, the
main identity we use from the general formula (\ref{eq:genSylv})
will be the case $m=3$, leading to the different
recurrence relations for (\ref{eq:Ppol}) and (\ref{eq:Qpol}).
So considering $m=3$ we obtain from (\ref{eq:genSylv}) the following 3-row/column Sylvester type identity:
\begin{eqnarray}
&& \left|
\begin{picture}(59,32)
\end{picture}
\right|\times
\left|
\begin{color}{red}
\begin{picture}(59,32)
\put(6,-26){\rule{.3mm}{2cm}}\put(11,-26){\rule{.3mm}{2cm}}\put(53,-26){\rule{.3mm}{2cm}}
\put(1,-21){\rule{2cm}{.3mm}}\put(1,21){\rule{2cm}{.3mm}}\put(1,26){\rule{2cm}{.3mm}}
\end{picture}
\end{color}
\right|
 =  \left|
\begin{color}{red}
\begin{picture}(59,32)
\put(6,-26){\rule{.3mm}{2cm}}\put(11,-26){\rule{.3mm}{2cm}}\put(1,21){\rule{2cm}{.3mm}}\put(1,-21){\rule{2cm}{.3mm}}
\end{picture}
\end{color}
\right|\times\left|
\begin{color}{red}
\begin{picture}(59,32)
\put(1,26){\rule{2cm}{.3mm}}\put(53,-26){\rule{.3mm}{2cm}}
\end{picture}
\end{color}
\right| \nn \\
 &-&\left|
\begin{color}{red}
\begin{picture}(59,32)
\put(6,-26){\rule{.3mm}{2cm}}\put(11,-26){\rule{.3mm}{2cm}}\put(1,-21){\rule{2cm}{.3mm}}\put(1,26){\rule{2cm}{.3mm}}
\end{picture}
\end{color}
\right|\times\left|
\begin{color}{red}
\begin{picture}(59,32)
\put(53,-26){\rule{.3mm}{2cm}}\put(1,21){\rule{2cm}{.3mm}}
\end{picture}
\end{color}
\right|
 +  \left|
\begin{color}{red}
\begin{picture}(59,32)
\put(6,-26){\rule{.3mm}{2cm}}\put(11,-26){\rule{.3mm}{2cm}}\put(1,21){\rule{2cm}{.3mm}}\put(1,26){\rule{2cm}{.3mm}}
\end{picture}
\end{color}
\right|\times\left|
\begin{color}{red}
\begin{picture}(59,32)
\put(53,-26){\rule{.3mm}{2cm}}\put(1,-21){\rule{2cm}{.3mm}}
\end{picture}
\end{color}
\right| , \nn \\
&& \quad\label{eq:3Sylv}
\end{eqnarray}
which is obtained from the expansion of (\ref{eq:genSylv}) for $m=3$ after recombination of terms using the earlier 2-row/column
Sylvester identity (\ref{eq:othSylv}).

\newpage
\section{Intermediate determinants}\label{appendix2}
\setcounter{equation}{0}
\def\theequation{B.\arabic{equation}}

Previously we have chosen to apply a 3 row/column Sylvester identity to the $P_{k}^{(l)}$ and $Q_{k}^{(l)}$ polynomials since this identity (\ref{eq:3Sylv}) does not introduce new determinants.  However for some 2 row/column Sylvester identities it is possible to control new determinants, which can be removed to still give equations in term $P_{k}^{(l)}$ and $Q_{k}^{(l)}$ only.  These {\it `` intermediate''} determinants are introduced here.

\subsection{The linear $P_{k}^{(l)}$ equation}\label{appendix2a}

By considering the $P_{k}^{(l)}$ with the first column and penultimate row removed we get the {\it intermediate} quantity:
\begin{equation}\label{eq:Tpol}
T^{(l)}_{k-1}(x,y)\equiv \left|\begin{array}{ccccc}
\langle\mbe_l,\mbe_2\rangle &\langle\mbe_l,\mbe_3\rangle & \cdots &\cdots &\langle\mbe_l,\mbe_k\rangle \\
\langle\mbe_{l+1},\mbe_2\rangle &\langle\mbe_{l+1},\mbe_3\rangle & \cdots &\cdots &\langle\mbe_{l+1},\mbe_k\rangle \\
\vdots  & \vdots  &   &  & \vdots \\
\vdots  & \vdots  &   &  & \vdots \\
\langle\mbe_{l+k-3},\mbe_2\rangle &\langle\mbe_{l+k-3},\mbe_3\rangle & \cdots &\cdots &\langle\mbe_{l+k-3},\mbe_k\rangle \\
\mbe_2  &\mbe_3  & \cdots &\cdots &\mbe_k
\end{array}\right|\big/ \Pi^{(l)}_{k-2}\quad ,
\end{equation}
together with a corresponding Hankel determinant:
\begin{equation}\label{eq:THankel}
\Pi^{(l)}_{k-1}\equiv\left|
\begin{array}{ccccc}
\langle\mbe_l,\mbe_2\rangle &\langle\mbe_l,\mbe_3\rangle & \cdots &\cdots &\langle\mbe_l,\mbe_k\rangle \\
\langle\mbe_{l+1},\mbe_2\rangle &\langle\mbe_{l+1},\mbe_3\rangle & \cdots &\cdots &\langle\mbe_{l+1},\mbe_k\rangle \\
\vdots  & \vdots  &   &  & \vdots \\
\vdots  & \vdots  &   &  & \vdots \\
\langle\mbe_{l+k-2},\mbe_2\rangle &\langle\mbe_{l+k-2},\mbe_3\rangle & \cdots &\cdots &\langle\mbe_{l+k-2},\mbe_k\rangle
\end{array}\right|\  .
\end{equation}
Using the usual Sylvester identity we can now derive the following two equations
\bse \label{eq:PTrels} \begin{eqnarray}
\left|
\begin{color}{red}
\begin{picture}(59,32)
\put(6,-26){\rule{.2mm}{2cm}}\put(53,-26){\rule{.2mm}{2cm}}
\put(1,-21){\rule{2cm}{.2mm}}\put(1,26){\rule{2cm}{.2mm}}
\begin{color}{black}
\put(20,2){\Large{$P_{k}^{(l)}$}}
\end{color}
\end{picture}
\end{color}
\right|
 & \Rightarrow &  P_k^{(l)}= T_{k-1}^{(l+1)} -
\frac{\Delta^{(l+1)}_{k-2}\Pi_{k-1}^{(l)}}{\Delta^{(l)}_{k-1}\Pi_{k-2}^{(l+1)}} P_{k-1}^{(l+1)}\  \label{eq:PTrelsa}, \\
\left|
\begin{color}{red}
\begin{picture}(59,32)
\put(6,-26){\rule{.2mm}{2cm}}\put(53,-26){\rule{.2mm}{2cm}}
\put(1,-21){\rule{2cm}{.2mm}}\put(1,-16){\rule{2cm}{.2mm}}
\begin{color}{black}
\put(20,2){\Large{$P_{k}^{(l)}$}}
\end{color}
\end{picture}
\end{color}
\right|
 & \Rightarrow &  P_{k}^{(l)}= T_{k-1}^{(l)} -
\frac{\Delta^{(l)}_{k-2}\Pi_{k-1}^{(l)}}{\Delta^{(l)}_{k-1}\Pi_{k-2}^{(l)}} P_{k-1}^{(l)}\  \label{eq:PTrelsb}.
\end{eqnarray}\ese
Eliminating the $T_k^{(l)}$ polynomials in favor of the $P_k^{(l)}$ polynomials and using a Hankel identity
\begin{equation}
\left|
\begin{color}{red}
\begin{picture}(59,32)
\put(6,-5){\rule{.2mm}{0.5cm}}\put(12,-26){\rule{.2mm}{2cm}}\put(53,-26){\rule{.2mm}{2cm}}
\put(1,-21){\rule{2cm}{.2mm}}\put(1,26){\rule{2cm}{.2mm}}
\begin{color}{black}
\put(20,2){\Large{$\Pi_{k}^{(l)}$}}
\end{color}
\end{picture}
\end{color}
\right|
\Rightarrow   \Pi_{k}^{(l)}\Delta_{k-2}^{(l+3)}  =  \Pi_{k-1}^{(l)}\Delta_{k-1}^{(l+3)}-\Pi_{k-1}^{(l+1)}\Delta_{k-1}^{(l+2)} \label{eq:Hankrel2}
\end{equation}
allows the derivation of the linear relation in $P_{k}^{(l)}$ (\ref{eq:PPrel}).

\subsection{The bilinear relation in $\Delta^{(l)}_{k},\Theta^{(l)}_{k}$}\label{appendix2b}
It is possible to combine some three-term bilinear Hankel relations, to give a three-term relation in terms of $\Delta^{(l)}_{k},\Theta^{(l)}_{k}$.  This particular identity is achieved with the introduction of a new {\it intermediate} determinant $\Xi$ which is essentially $\Theta$ with the first column and last row removed.
\bse\label{eq:Hankrel35}\begin{eqnarray}
\left|
\begin{color}{red}
\begin{picture}(59,32)
\put(6,-26){\rule{.2mm}{2cm}}\put(53,-26){\rule{.2mm}{2cm}}
\put(1,21){\rule{2cm}{.2mm}}\put(1,26){\rule{2cm}{.2mm}}
\begin{color}{black}
\put(20,2){\Large{$\Delta_{k}^{(l)}$}}
\end{color}
\end{picture}
\end{color}
\right|
&  \Rightarrow  &  \Delta_k^{(l)}\Pi_{k-2}^{(l+2)}= \Theta_{k-1}^{(l)}\Pi_{k-1}^{(l+1)}- \Delta_{k-1}^{(l+1)}\Xi_{k-1}^{(l)} \label{eq:Hankrel3} \\
\left|
\begin{color}{red}
\begin{picture}(59,32)
\put(6,-26){\rule{.2mm}{2cm}}\put(53,-26){\rule{.2mm}{2cm}}
\put(1,-21){\rule{2cm}{.2mm}}\put(1,26){\rule{2cm}{.2mm}}
\begin{color}{black}
\put(20,2){\Large{$\Theta_{k}^{(l)}$}}
\end{color}
\end{picture}
\end{color}
\right|
&  \Rightarrow  & \Theta_{k}^{(l)}\Pi_{k-2}^{(l+2)} = \Theta_{k-1}^{(l)}\Pi_{k-1}^{(l+2)} - \Delta_{k-1}^{(l+2)}\Xi_{k-1}^{(l)} \label{eq:Hankrel5}
\end{eqnarray}\ese
where
\begin{equation}\label{eq:GHankel}
\Xi^{(l)}_{k-1}\equiv\left|
\begin{array}{ccccc}
\langle\mbe_l,\mbe_2\rangle &\langle\mbe_l,\mbe_3\rangle & \cdots &\cdots &\langle\mbe_l,\mbe_k\rangle \\
\langle\mbe_{l+2},\mbe_2\rangle &\langle\mbe_{l+2},\mbe_3\rangle & \cdots &\cdots &\langle\mbe_{l+2},\mbe_k\rangle \\
\vdots  & \vdots  &   &  & \vdots \\
\vdots  & \vdots  &   &  & \vdots \\
\langle\mbe_{l+k-1},\mbe_2\rangle &\langle\mbe_{l+k-1},\mbe_3\rangle & \cdots &\cdots &\langle\mbe_{l+k-1},\mbe_k\rangle
\end{array}\right|\  .
\end{equation}

$\Xi$ is then eliminated in the combination of the two bilinear relations (\ref{eq:Hankrel35}).
\begin{equation}\label{eq:C4}
\Delta_{k-1}^{(l+1)}(\Theta_{k-1}^{(l)}\Pi_{k-1}^{(l+2)}-\Theta_{k}^{(l)}\Pi_{k-2}^{(l+2)})=\Delta_{k-1}^{(l+2)}(\Theta_{k-1}^{(l)}\Pi_{k-1}^{(l+1)}-\Delta_{k}^{(l)}\Pi_{k-2}^{(l+2)})
\end{equation}
This equation can be expanded (in the third term) using a third bilinear equation
\begin{equation}
\left|
\begin{color}{red}
\begin{picture}(59,32)
\put(6,-26){\rule{.2mm}{2cm}}\put(53,-26){\rule{.2mm}{2cm}}
\put(1,-21){\rule{2cm}{.2mm}}\put(1,26){\rule{2cm}{.2mm}}
\begin{color}{black}
\put(20,2){\Large{$\Delta_{k}^{(l)}$}}
\end{color}
\end{picture}
\end{color}
\right|
\Rightarrow   \Delta_k^{(l)}\Pi_{k-2}^{(l+1)}= \Delta_{k-1}^{(l)}\Pi_{k-1}^{(l+1)}- \Delta_{k-1}^{(l+1)}\Pi_{k-1}^{(l)}\ .  \label{eq:Hankrel1}
\end{equation}
Thus we are left with an equation in terms of $\Delta$ and $\Theta$ only.
\bea
\Delta_{k-1}^{(l+1)}(\Theta_{k-1}^{(l)}\Pi_{k-1}^{(l+2)}-\Theta_{k}^{(l)}\Pi_{k-2}^{(l+2)})&=&\Theta_{k-1}^{(l)}(\Delta_{k-1}^{(l+1)}\Pi_{k-1}^{(l+2)}-\Delta_{k}^{(l+1)}\Pi_{k-2}^{(l+2)})\nn  \\
&&-\Delta_{k-1}^{(l+2)}\Delta_{k}^{(l)}\Pi_{k-2}^{(l+2)} \nn
\eea
After we have canceled the necessary terms, we are left with (\ref{eq:Hankrel7}).

\subsection{Relations in $P_{k}^{(l)}$ and $Q_{k}^{(l)}$}\label{appendix2c}
For relations involving the Hankel determinants $P_{k}^{(l)}$ and $Q_{k}^{(l)}$ we can derive a recurrence relation $xQ_{k}^{(l)}$ and two linear relations.
By implementing a cutting of rows and columns on the matrix for $P_k^{(l)}$ according to:
{\tiny\begin{eqnarray*}
\left|
\begin{picture}(59,32)
\end{picture}
\right|&\times&
\left|
\begin{color}{red}
\begin{picture}(59,32)
\put(6,-26){\rule{.3mm}{2cm}}\put(9,-5){\rule{.2mm}{0.5cm}}\put(12,-26){\rule{.3mm}{2cm}}\put(53,-26){\rule{.3mm}{2cm}}
\put(1,-21){\rule{2cm}{.3mm}}\put(1,-16){\rule{2cm}{.3mm}}\put(1,21){\rule{2cm}{.3mm}}
\end{picture}
\end{color}
\right|
 =  \left|
\begin{color}{red}
\begin{picture}(59,32)
\put(6,-26){\rule{.3mm}{2cm}}\put(9,-5){\rule{.2mm}{0.5cm}}\put(12,-26){\rule{.3mm}{2cm}}\put(1,-16){\rule{2cm}{.3mm}}\put(1,21){\rule{2cm}{.3mm}}
\end{picture}
\end{color}
\right|\times\left|
\begin{color}{red}
\begin{picture}(59,32)
\put(53,-26){\rule{.3mm}{2cm}}\put(1,-21){\rule{2cm}{.3mm}}
\end{picture}
\end{color}
\right| \\
&-&\left|
\begin{color}{red}
\begin{picture}(59,32)
\put(6,-26){\rule{.3mm}{2cm}}\put(9,-5){\rule{.2mm}{0.5cm}}\put(12,-26){\rule{.3mm}{2cm}}\put(1,-21){\rule{2cm}{.3mm}}\put(1,21){\rule{2cm}{.3mm}}
\end{picture}
\end{color}
\right|\times\left|
\begin{color}{red}
\begin{picture}(59,32)
\put(53,-26){\rule{.3mm}{2cm}}\put(1,-16){\rule{2cm}{.3mm}}
\end{picture}
\end{color}
\right|
 +  \left|
\begin{color}{red}
\begin{picture}(59,32)
\put(6,-26){\rule{.3mm}{2cm}}\put(9,-5){\rule{.2mm}{0.5cm}}\put(12,-26){\rule{.3mm}{2cm}}\put(1,-16){\rule{2cm}{.3mm}}\put(1,-21){\rule{2cm}{.3mm}}
\end{picture}
\end{color}
\right|\times\left|
\begin{color}{red}
\begin{picture}(59,32)
\put(1,21){\rule{2cm}{.3mm}}\put(53,-26){\rule{.3mm}{2cm}}
\end{picture}
\end{color}
\right|
\end{eqnarray*}}
and applying the 3-row/column Sylvester identity in that situation, we obtain:
\begin{equation}\label{eq:xQrec}
P_k^{(l)}=x Q_{k-2}^{(l+2)} -~\frac{\Delta_{k-2}^{(l)} \Theta^{(l+2)}_{k-2}}{\Delta_{k-1}^{(l)}\Theta_{k-3}^{(l+2)}}~P_{k-1}^{(l)}
+~\frac{\Theta_{k-2}^{(l)} \Delta^{(l+2)}_{k-2}}{\Delta_{k-1}^{(l)}\Theta_{k-3}^{(l+2)}}~Q_{k-1}^{(l)}\quad ,
\quad l\neq 0,1\  .
\end{equation}
For the linear relations we first obtain
\begin{equation}
\left|
\begin{color}{red}
\begin{picture}(59,32)
\put(6,-26){\rule{.2mm}{2cm}}\put(53,-26){\rule{.2mm}{2cm}}
\put(1,-21){\rule{2cm}{.2mm}}\put(1,26){\rule{2cm}{.2mm}}
\begin{color}{black}
\put(20,2){\Large{$Q_{k}^{(l)}$}}
\end{color}
\end{picture}
\end{color}
\right|
\Rightarrow  \qquad Q_k^{(l)}= T_{k-1}^{(l+2)} -
\frac{\Delta^{(l+2)}_{k-2}\Gamma_{k-1}^{(l)}}{\Theta^{(l)}_{k-1}\Pi_{k-2}^{(l+2)}} P_{k-1}^{(l+2)}  ,
\end{equation}
which by eliminating the $T_k^{(l)}$ using (\ref{eq:PTrelsa}) together with the Hankel identity (\ref{eq:Hankrel3}) leads to
\begin{equation}\label{eq:QPPrel}
Q_k^{(l)}= P_k^{(l+1)} +
\frac{\Delta^{(l)}_k\Delta_{k-2}^{(l+2)}}{\Theta^{(l)}_{k-1}\Delta_{k-1}^{(l+1)}} P_{k-1}^{(l+2)}\  .
\end{equation}
This three term equation in terms of $Q_{k}^{(l)}$ and $P_{k}^{(l)}$ is similar to (\ref{eq:PPrel}).  To find another of this type of equation we must first introduce the {\it intermediate} polynomials $S^{(l)}_{k-1}$ :
\begin{equation}\label{eq:Spol}
S^{(l)}_{k-1}(x,y)\equiv \left|\begin{array}{ccccc}
\langle\mbe_l,\mbe_2\rangle &\langle\mbe_l,\mbe_3\rangle & \cdots &\cdots &\langle\mbe_l,\mbe_k\rangle \\
\langle\mbe_{l+2},\mbe_2\rangle &\langle\mbe_{l+2},\mbe_3\rangle & \cdots &\cdots &\langle\mbe_{l+2},\mbe_k\rangle \\
\vdots  & \vdots  &   &  & \vdots \\
\vdots  & \vdots  &   &  & \vdots \\
\langle\mbe_{l+k-2},\mbe_2\rangle &\langle\mbe_{l+k-2},\mbe_3\rangle & \cdots &\cdots &\langle\mbe_{l+k-2},\mbe_k\rangle \\
\mbe_2  &\mbe_3  & \cdots &\cdots &\mbe_k
\end{array}\right|\big/ \Gamma^{(l)}_{k-2}\quad .
\end{equation}
Then applying the following Sylvester identities to $P_{k}^{(l)}$ and $Q_{k}^{(l)}$, we subsequently obtain the relations:
\bse\begin{eqnarray}
\left|
\begin{color}{red}
\begin{picture}(59,32)
\put(6,-26){\rule{.2mm}{2cm}}\put(53,-26){\rule{.2mm}{2cm}}
\put(1,-21){\rule{2cm}{.2mm}}\put(1,-16){\rule{2cm}{.2mm}}
\begin{color}{black}
\put(20,2){\Large{$Q_{k}^{(l)}$}}
\end{color}
\end{picture}
\end{color}
\right|
 & \Rightarrow  &  Q_k^{(l)} = S_{k-1}^{(l)}-\frac{\Gamma_{k-1}^{(l)}\Theta_{k-2}^{(l)}}
{\Gamma_{k-2}^{(l)}\Theta_{k-1}^{(l)}} Q_{k-1}^{(l)}\  , \\
\left|
\begin{color}{red}
\begin{picture}(59,32)
\put(6,-26){\rule{.2mm}{2cm}}\put(53,-26){\rule{.2mm}{2cm}}
\put(1,-21){\rule{2cm}{.2mm}}\put(1,21){\rule{2cm}{.2mm}}
\begin{color}{black}
\put(20,2){\Large{$P_{k}^{(l)}$}}
\end{color}
\end{picture}
\end{color}
\right|
& \Rightarrow &  P_k^{(l)} = S_{k-1}^{(l)}-\frac{\Pi_{k-1}^{(l)}\Theta_{k-2}^{(l)}}
{\Gamma_{k-2}^{(l)}\Delta_{k-1}^{(l)}} Q_{k-1}^{(l)}\  .
\end{eqnarray}\ese
We eliminate  $S_{k-1}^{(l)}$ and making use of the Hankel determinant identity
\begin{equation}
\left|
\begin{color}{red}
\begin{picture}(59,32)
\put(6,-26){\rule{.2mm}{2cm}}\put(53,-26){\rule{.2mm}{2cm}}
\put(1,-21){\rule{2cm}{.2mm}}\put(1,21){\rule{2cm}{.2mm}}
\begin{color}{black}
\put(20,2){\Large{$\Delta_{k}^{(l)}$}}
\end{color}
\end{picture}
\end{color}
\right|
\Rightarrow   \Delta_k^{(l)}\Xi_{k-2}^{(l)}= \Delta_{k-1}^{(l)}\Xi_{k-1}^{(l)}- \Theta_{k-1}^{(l)}\Pi_{k-1}^{(l)} \label{eq:Hankrel4}
\end{equation}
which leaves:
\begin{equation}\label{eq:QPQrel}
Q_k^{(l)}=P_k^{(l)}-\frac{\Delta_k^{(l)}\Theta_{k-2}^{(l)}}
{\Delta_{k-1}^{(l)}\Theta_{k-1}^{(l)}} Q_{k-1}^{(l)}  \ .
\end{equation}

\newpage
\section{$\s$-Periodic initial value problems for system (\ref{eq:HS2})}\label{appendix4}
\setcounter{equation}{0}
\def\theequation{C.\arabic{equation}}

Proposition \ref{PBH} is obtained by considering the ranges of the $n$-values for the equations with periods taken
in the remaining regions. When $\s\in R^2$ (see Figure \ref{DRBH}) we get the following reduction.

\begin{figure}[h]
\setlength{\unitlength}{8mm}
\begin{center}
\begin{picture}(8,2.5)(0,0)
\path(0,0)(8,0)
\multiput(0,0)(2,0){5}{\path(0,-.1)(0,.1)}
\thicklines
\path(0,1)(8,1)
\path(0,2)(6,2)
\multiput(0,1)(0,1){2}{\circle*{.15}}
\put(2,2){\circle*{.15}}
\put(4,1){\circle*{.15}}
\put(6,2){\circle*{.15}}
\put(8,1){\circle*{.15}}
\put(-.1,.2){$0$}
\put(1.9,.2){$b$}
\put(3.4,.2){$2a-b$}
\put(5.8,.2){$2a$}
\put(7.4,.2){$4a-b$}
\put(-.3,2.1){$\Thetar$}
\put(-.4,1.1){$\Deltas\ \Thetar$}
\put(1.6,2.1){$\Deltas\ \Thetar$}
\put(3.7,1.1){$\Thetar$}
\put(5.7,2.1){$\Deltas$}
\put(7.7,1.1){$\Deltas$}
\end{picture}
\captionof{figure}{$\s$-Reduction for system (\ref{eq:HS2}), with $\s\in R^2$}
\end{center}
\end{figure}
So we first need to calculate the values of $\Thetar$ for $b\leq n \leq 2a-b$. This can be done using equation
(\ref{eq:Hankrel}) because the values of $\Deltas$ at $n$ smaller than $2a+(2a-b-b)=4a-2b$ are given initially.
Then the values of $\Deltas$ at $n=4a-b$ are determined by (\ref{eq:Hankrel7}). When $\s\in R^3$ we get

\begin{figure}[h!]
\setlength{\unitlength}{8mm}
\begin{center}
\begin{picture}(10,2.5)(0,0)
\path(0,0)(8,0)
\dottedline{.15}(8,0)(10,0)
\multiput(0,0)(2,0){6}{\path(0,-.1)(0,.1)}
\thicklines
\path(0,1)(10,1)
\path(0,2)(8,2)
\multiput(0,1)(0,1){2}{\circle*{.15}}
\put(2,1){\circle*{.15}}
\put(4,2){\circle*{.15}}
\put(6,2){\circle*{.15}}
\put(8,2){\circle*{.15}}
\put(10,1){\circle*{.15}}
\put(-.1,.2){$0$}
\put(1.4,.2){$2a-b$}
\put(3.9,.2){$a$}
\put(5.9,.2){$b$}
\put(7.6,.2){$a+b$}
\put(9.4,.2){$4a-b$}
\put(-.3,2.1){$\Thetar$}
\put(-.3,1.1){$\Deltas\ \Thetar$}
\put(1.7,1.1){$\Thetar$}
\put(3.7,2.1){$\Deltas$}
\put(5.7,2.1){$\Thetar$}
\put(7.7,2.1){$\Deltas$}
\put(9.7,1.1){$\Deltas$}
\end{picture}
\captionof{figure}{\label{r3} $\s$-Reduction for the system (\ref{eq:HS2}), with $\s\in R^3$}
\end{center}
\end{figure}
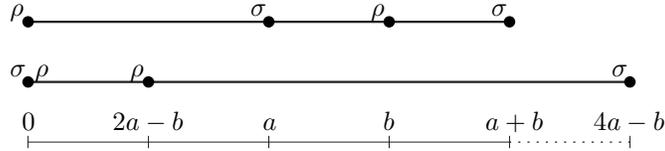

In Figure \ref{r3} the dashed line indicates that $a+b$ can either be to the left, or to the right
of $4a-b$. We have $a+b<4a-b$ in the region $3a>2b>2a$. Here it does not matter whether one first calculates
the values of $\Thetar$ at $n=b$, using equation (\ref{eq:Hankrel}), or the values of $\Deltas$ at
$n=4a-b$, using equation (\ref{eq:Hankrel7}). In the region $3a\leq 2b < 4a$ we have
$a+b\geq 4a-b$. Here we need first calculate the values of $\Deltas$ for $4a-b\leq n \leq a+b$,
using equation (\ref{eq:Hankrel7}). This can be done indeed, because the values of $\Thetar$
with $n\leq 2a-b+(a+b)-(4a-b)=b-a$ are given initially. Subsequently the values of $\Thetar$
at $n=b$ can be obtained.

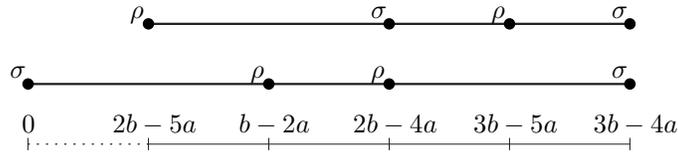
\begin{figure}[h!]
\setlength{\unitlength}{8mm}
\begin{center}
\begin{picture}(10,2.5)(0,0)
\path(2,0)(10,0)
\dottedline{.15}(0,0)(2,0)
\multiput(0,0)(2,0){6}{\path(0,-.1)(0,.1)}
\thicklines
\path(0,1)(10,1)
\path(2,2)(10,2)
\put(0,1){\circle*{.15}}
\put(2,2){\circle*{.15}}
\put(4,1){\circle*{.15}}
\put(6,1){\circle*{.15}}
\put(6,2){\circle*{.15}}
\put(8,2){\circle*{.15}}
\put(10,1){\circle*{.15}}
\put(10,2){\circle*{.15}}
\put(-.1,.2){$0$}
\put(1.4,.2){$2b-5a$}
\put(3.5,.2){$b-2a$}
\put(5.4,.2){$2b-4a$}
\put(7.4,.2){$3b-5a$}
\put(9.4,.2){$3b-4a$}
\put(-.3,1.1){$\Deltas$}
\put(1.7,2.1){$\Thetar$}
\put(3.7,1.1){$\Thetar$}
\put(5.7,1.1){$\Thetar$}
\put(5.7,2.1){$\Deltas$}
\put(7.7,2.1){$\Thetar$}
\put(9.7,1.1){$\Deltas$}
\put(9.7,2.1){$\Deltas$}
\end{picture}
\captionof{figure}{\label{r41} $\s$-Reduction for the system (\ref{eq:HS2}), with $s\in R^4$, $b \leq 3a$}
\end{center}
\end{figure}

When $\s\in R^4$ we also need to distinguish two cases, but here this leads to different initial value problems.
When $2a > b \leq 3a$ we can first calculate the values of $\Deltas$ at $n=3b-4a$, and then the values of $\Thetar$
at $n=3b-5a$, see Figure \ref{r41}.

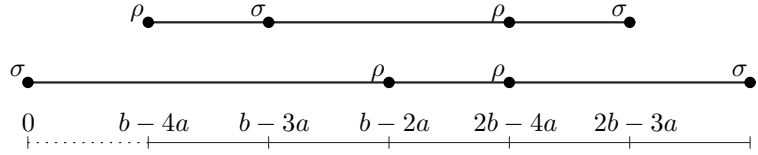
\begin{figure}[h!]
\setlength{\unitlength}{8mm}
\begin{center}
\begin{picture}(12,2.5)(0,0)
\path(2,0)(12,0)
\dottedline{.15}(0,0)(2,0)
\multiput(0,0)(2,0){7}{\path(0,-.1)(0,.1)}
\thicklines
\path(0,1)(12,1)
\path(2,2)(10,2)
\put(0,1){\circle*{.15}}
\put(2,2){\circle*{.15}}
\put(4,2){\circle*{.15}}
\put(6,1){\circle*{.15}}
\put(8,1){\circle*{.15}}
\put(8,2){\circle*{.15}}
\put(10,2){\circle*{.15}}
\put(12,1){\circle*{.15}}
\put(-.1,.2){$0$}
\put(1.5,.2){$b-4a$}
\put(3.5,.2){$b-3a$}
\put(5.5,.2){$b-2a$}
\put(7.4,.2){$2b-4a$}
\put(9.4,.2){$2b-3a$}
\put(19.4,.2){$3b-4a$}
\put(-.3,1.1){$\Deltas$}
\put(1.7,2.1){$\Thetar$}
\put(3.7,2.1){$\Deltas$}
\put(5.7,1.1){$\Thetar$}
\put(7.7,1.1){$\Thetar$}
\put(7.7,2.1){$\Thetar$}
\put(9.7,2.1){$\Deltas$}
\put(11.7,1.1){$\Deltas$}
\end{picture}
\captionof{figure}{\label{r42} $\s$-Reduction for the system (\ref{eq:HS2}), with $s\in R^4$, $b \geq 3a$}
\end{center}
\end{figure}

When $b \geq 3a$ we need first calculate the values of $\Thetar$ at $n=2b-4a$, and then the values of $\Deltas$ at $n=3b-4a$, see Figure \ref{r42}.

We leave it to the reader to check that the given initial value problems can also be updated in
the negative $n$-direction. In Appendix \ref{appendix6} we illustrate with an example that this is not necessarily the case,
and thus has to be verified separately.

\newpage
\section{$\s$-Periodic initial value problems for the QQD scheme}\label{appendix5}
\setcounter{equation}{0}
\def\theequation{D.\arabic{equation}}

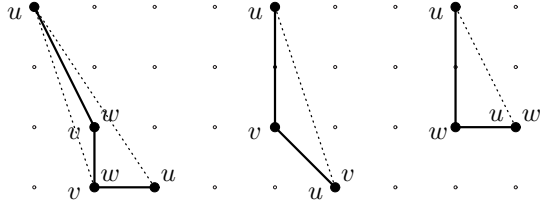
\begin{figure}[h!]
\setlength{\unitlength}{8mm}
\begin{picture}(8,3)(0,0)
\matrixput(0,0)(1,0){9}(0,1){4}{\circle{.05}}
\multiput(1,0)(0,1){2}{\circle*{.15}}
\multiput(2,0)(-2,3){2}{\circle*{.15}}

\multiput(5,0)(-1,1){2}{\circle*{.15}}
\put(4,3){\circle*{.15}}

\multiput(7,1)(1,0){2}{\circle*{.15}}
\put(7,3){\circle*{.15}}

\dottedline{.1}(0,3)(1,0)
\dottedline{.1}(0,3)(2,0)

\dottedline{.1}(4,3)(5,0)

\dottedline{.1}(7,3)(8,1)

\thicklines
\path(0,3)(1,1)(1,0)(2,0)
\put(-.45,2.8){$\A$}
\put(.55,.8){$\B$}
\put(1.1,1.1){$\C$}
\put(.55,-.2){$\B$}
\put(1.1,0.1){$\C$}
\put(2.1,.1){$\A$}

\path(4,3)(4,1)(5,0)
\put(3.55,2.8){$\A$}
\put(3.55,.8){$\B$}
\put(4.55,-.2){$\A$}
\put(5.1,.1){$\B$}

\path(7,3)(7,1)(8,1)
\put(6.55,2.8){$\A$}
\put(6.55,.8){$\C$}
\put(7.55,1.1){$\A$}
\put(8.1,1.1){$\C$}

\end{picture}
\caption{\label{FQQD} Graphical representation of the QQD scheme}
\end{figure}

The stencils of the QQD scheme (\ref{eq:QQD}) are depicted in Figure \ref{FQQD}.

To describe initial value problems for the QQD scheme we follow the following procedure.
We project the stencils of the three equations onto lines with directions in different
regions and translate them in such a way that, given the values of the fields for certain
ranges of $n$, each equation determines a value for an update of one of the fields.
The proof of Proposition consists of 5 pictures, one for each region.
For every picture one has to check the following:
\begin{enumerate}
\item that the ranges of $n$-values for $\A,\B,\C$ correspond to each equation in the system,
for any chosen direction $\s$ in the particular region, this includes checking the order of the
linear expressions in $a,b$.
\item that, given the initial values for $\A,\B,\C$, every equation can be used to update one of
the fields (in both directions).
\end{enumerate}

\begin{figure}[h!]
\setlength{\unitlength}{8mm}
\begin{center}
\begin{picture}(12,3.5)(0,0)
\path(0,0)(2,0)
\dottedline{.15}(2,0)(4,0)
\path(4,0)(8,0)
\dottedline{.15}(8,0)(10,0)
\path(10,0)(12,0)
\multiput(0,0)(2,0){7}{\path(0,-.1)(0,.1)}
\thicklines
\path(2,1)(8,1)
\path(0,2)(10,2)
\path(0,3)(12,3)
\multiput(0,2)(0,1){2}{\circle*{.15}}
\put(2,1){\circle*{.15}}
\put(4,2){\circle*{.15}}
\multiput(6,1)(0,2){2}{\circle*{.15}}
\put(8,1){\circle*{.15}}
\multiput(10,2)(0,1){2}{\circle*{.15}}
\put(12,3){\circle*{.15}}
\put(-.1,.2){$0$}
\put(1.9,.2){$b$}
\put(3.8,.2){$2a$}
\put(5.5,.2){$2a+b$}
\put(7.5,.2){$2a+2b$}
\put(9.5,.2){$3a+b$}
\put(11.5,.2){$3a+2b$}
\put(-.3,2.1){$\A$}
\put(-.3,3.1){$\A$}
\put(1.7,1.1){$\A$}
\put(3.7,2.1){$\B$}
\put(5.7,1.1){$\C$}
\put(5.7,3.1){$\B$}
\put(6.1,3.1){$\C$}
\put(7.7,1.1){$\A$}
\put(8.1,1.1){$\C$}
\put(9.7,2.1){$\A$}
\put(10.1,2.1){$\B$}
\put(9.7,3.1){$\B$}
\put(10.1,3.1){$\C$}
\put(11.7,3.1){$\A$}
\end{picture}
\caption{\label{qqdr1} $\s$-Reduction of the QQD scheme, $\s\in R^1$}
\end{center}
\end{figure}
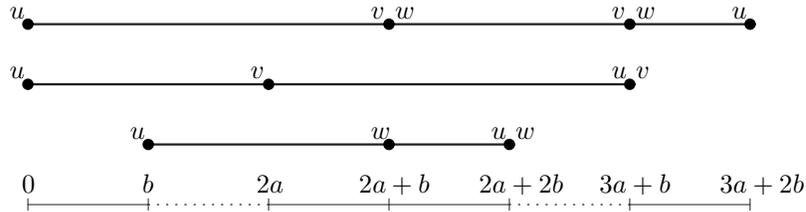
We will perform check (1) and check (2, positive direction) for the first picture, and check (2, positive direction) for the other
pictures. This enables one to perform $\s$-reduction for any given $\s\in \cup_i R^i$. The rest of the proof is left to the reader.

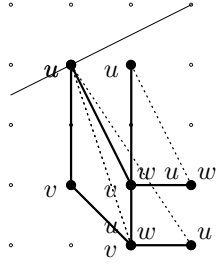
\begin{figure}[h!]
\setlength{\unitlength}{8mm}
\begin{picture}(3,4)(0,0)
\matrixput(0,0)(1,0){4}(0,1){5}{\circle{.05}}

\multiput(2,0)(0,1){2}{\circle*{.15}}
\multiput(3,0)(-2,3){2}{\circle*{.15}}

\multiput(2,0)(-1,1){2}{\circle*{.15}}
\put(1,3){\circle*{.15}}

\multiput(2,1)(1,0){2}{\circle*{.15}}
\put(2,3){\circle*{.15}}

\dottedline{.1}(1,3)(2,0)
\dottedline{.1}(1,3)(3,0)

\dottedline{.1}(1,3)(2,0)

\dottedline{.1}(2,3)(3,1)

\path(0,2.5)(3,4)

\thicklines
\path(1,3)(2,1)(2,0)(3,0)
\put(.55,2.8){$\A$}
\put(1.55,.8){$\B$}
\put(2.1,1.1){$\C$}
\put(1.55,-.2){$\B$}
\put(2.1,0.1){$\C$}
\put(3.1,.1){$\A$}

\path(1,3)(1,1)(2,0)
\put(.55,2.8){$\A$}
\put(.55,.8){$\B$}
\put(1.55,.2){$\A$}

\path(2,3)(2,1)(3,1)
\put(1.55,2.8){$\A$}
\put(2.55,1.1){$\A$}
\put(3.1,1.1){$\C$}

\end{picture}
\caption{\label{FQQDO} The equations of the QQD scheme on top of each other, and a line with direction $(2,1)\in R^1$.}
\end{figure}

When $\epsilon=1$, as in region $R^1$, with one step to the right (downwards) $n$ increases by $b$ (by $a$).
Let $n$ be equal to $0$ on a line with direction $\s\in R^1$ through the first point $\A$ in the first
equation of the system (the upper-left point in Figure \ref{FQQD}). We let the upper-left point from
the second equation coincide with the one of the first equation and we let the upper-left point from
the third equation be one step to the right, so that it is at distance $b$ from the others, as in Figure
\ref{FQQDO}.
Moving the line over the Figure \ref{FQQDO} in downward direction, depending on the direction $\s$, the line will either
first cross this point at distance $b$, or the point in the second equation at distance $2a$. This
is indicated in Figure \ref{qqdr1} by the dashed line between $b$ and $2a$. Next the line moves over the point
with distance $2a+b$, and then it depends on $\s$ whether $2a+2b$ or $3a+b$ is encountered first. This is
again indicated by a dashed line. Thus, Figure \ref{qqdr1} represents the ranges of the $n$-values of the
three fields of the three equations in the QQD scheme and their relative positions, for $\s\in R^1$.

Next, suppose initially we are given the values of $\A$ at $0\leq n < 3a+2b$, of $\B$ at $2a \leq n < 3a+b$, and of
$\C$ at $2a+b \leq n < 2a+2b$. Then, for all $\s\in R^1$, we can use the second equation to determine the value
of $\B$ at $n=3a+b$. If $a<b$, then we know all values of $\C$ at $2a+b \leq n < 3a+b$, and hence we can use the
first equation to calculate the value of $\A$ at $n=3a+2b$. Then, we can calculate the value of $\C$ at $n=2a+2b$.
However, if $a\geq b$, then we first have to use the last equation to calculate all values of $\C$ at $2a+2b\leq n
\leq 3a+b$, before we can use the first equation to calculate the value of $\A$ at $n=3a+2b$. In both cases, the
initial value problem is well-posed.

\begin{figure}[h!]
\setlength{\unitlength}{8mm}
\begin{center}
\begin{picture}(10,3.5)(0,0)
\path(0,0)(10,0)
\multiput(0,0)(2,0){6}{\path(0,-.1)(0,.1)}
\thicklines
\path(0,1)(6,1)
\path(2,2)(10,2)
\path(0,3)(8,3)
\multiput(0,1)(0,2){2}{\circle*{.15}}
\multiput(2,1)(0,1){3}{\circle*{.15}}
\multiput(4,2)(0,1){2}{\circle*{.15}}
\put(6,1){\circle*{.15}}
\put(8,3){\circle*{.15}}
\put(10,2){\circle*{.15}}
\put(-.1,.2){$0$}
\put(1.9,.2){$b$}
\put(3.9,.2){$a$}
\put(5.8,.2){$2a$}
\put(7.5,.2){$3a-b$}
\put(9.8,.2){$3a$}
\put(-.3,3.1){$\B$}
\put(.1,3.1){$\C$}
\put(-.3,1.1){$\C$}
\put(1.7,2.1){$\A$}
\put(1.7,3.1){$\A$}
\put(2.1,2.1){$\B$}
\put(1.7,1.1){$\A$}
\put(2.1,1.1){$\C$}
\put(3.7,2.1){$\B$}
\put(3.7,3.1){$\B$}
\put(4.1,3.1){$\C$}
\put(5.7,1.1){$\A$}
\put(7.7,3.1){$\A$}
\put(9.7,2.1){$\A$}
\end{picture}
\caption{\label{qqdr2} $\s$-Reduction of the QQD scheme, $\s\in R^2$}
\end{center}
\end{figure}
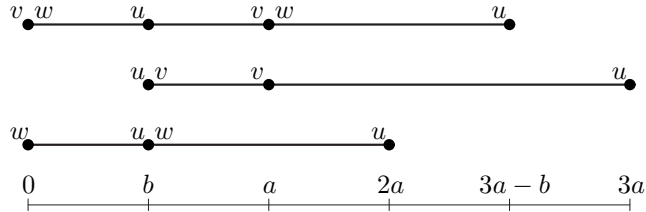

In $R^2$ we first use the third equation to determine the values of $\C$ at $b\leq n\leq a$, then the
first equation to determine the values of $\B$ at $n=a$, and finally the second equation to
determine the values of $\A$ at $n=3a$.

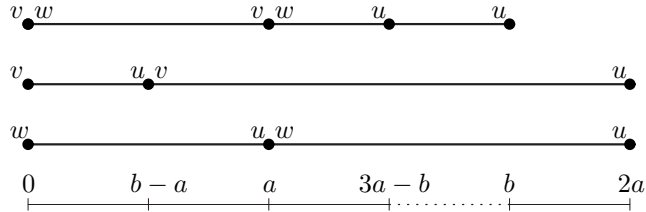
\begin{figure}[h!]
\setlength{\unitlength}{8mm}
\begin{center}
\begin{picture}(10,3.5)(0,0)
\path(0,0)(6,0)
\dottedline{.15}(6,0)(8,0)
\path(8,0)(10,0)
\multiput(0,0)(2,0){6}{\path(0,-.1)(0,.1)}
\thicklines
\path(0,1)(10,1)
\path(0,2)(10,2)
\path(0,3)(8,3)
\multiput(0,1)(0,1){3}{\circle*{.15}}
\put(0,3){\circle*{.15}}
\put(2,2){\circle*{.15}}
\multiput(4,1)(0,2){2}{\circle*{.15}}
\multiput(6,3)(2,0){2}{\circle*{.15}}
\multiput(10,1)(0,1){2}{\circle*{.15}}
\put(-.1,.2){$0$}
\put(1.7,.2){$b-a$}
\put(3.9,.2){$a$}
\put(5.5,.2){$3a-b$}
\put(7.9,.2){$b$}
\put(9.8,.2){$2a$}
\put(-.3,3.1){$\B$}
\put(.1,3.1){$\C$}
\put(-.3,2.1){$\B$}
\put(-.3,1.1){$\C$}
\put(1.7,2.1){$\A$}
\put(2.1,2.1){$\B$}
\put(3.7,1.1){$\A$}
\put(4.1,3.1){$\C$}
\put(4.1,1.1){$\C$}
\put(3.7,3.1){$\B$}
\multiput(5.7,3.1)(2,0){2}{$\A$}
\put(9.7,1.1){$\A$}
\put(9.7,2.1){$\A$}
\end{picture}
\caption{\label{qqdr3} $\s$-Reduction of the QQD scheme, $\s\in R^3$}
\end{center}
\end{figure}

In $R^3$ the values of $\A$ at $n=2a$ and the values of $\B$ at $n=a$ can be determined independently, using the
second and first equation, respectively. After having determined the values of $\A$ at $n=2a$ we are able to find
the $\C$ at $n=b$, using the third equation.

\begin{figure}[h!]
\setlength{\unitlength}{8mm}
\begin{center}
\begin{picture}(10,3.5)(0,0)
\path(0,0)(10,0)
\multiput(0,0)(2,0){6}{\path(0,-.1)(0,.1)}
\thicklines
\path(0,1)(10,1)
\path(0,2)(8,2)
\path(0,3)(10,3)
\multiput(0,1)(0,1){3}{\circle*{.15}}
\put(2,3){\circle*{.15}}
\put(4,3){\circle*{.15}}
\put(6,2){\circle*{.15}}
\multiput(8,1)(0,1){2}{\circle*{.15}}
\multiput(10,1)(0,2){2}{\circle*{.15}}
\put(-.1,.2){$0$}
\put(1.5,.2){$3a-b$}
\put(3.9,.2){$a$}
\put(5.7,.2){$b-a$}
\put(7.8,.2){$2a$}
\put(9.9,.2){$b$}
\put(-.3,3.1){$\B$}
\put(-.3,2.1){$\B$}
\put(-.3,1.1){$\C$}
\put(.1,3.1){$\C$}
\put(1.7,3.1){$\A$}
\put(3.7,3.1){$\B$}
\put(4.1,3.1){$\C$}
\put(5.7,2.1){$\A$}
\put(6.1,2.1){$\B$}
\put(7.7,1.1){$\A$}
\put(7.7,2.1){$\A$}
\put(9.7,3.1){$\A$}
\put(9.7,1.1){$\A$}
\put(10.1,1.1){$\C$}
\end{picture}
\caption{\label{qqdr4} $\s$-Reduction of the QQD scheme, $\s\in R^4$}
\end{center}
\end{figure}
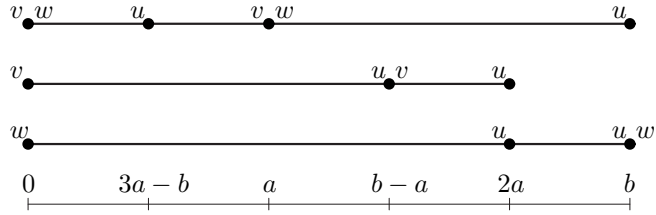

In $R^4$ the values of $\A$ at $n=b$ and the values of $\B$ at $n=b-a$ can be determined independently, using the
first and second equation, respectively. After having determined the values of $\A$ at $n=b$ we are able to find
the $\C$ at $n=b$, using the third equation.

\begin{figure}[h!]
\setlength{\unitlength}{8mm}
\begin{center}
\begin{picture}(12,3.5)(0,0)
\path(0,0)(6,0)
\dottedline{.15}(6,0)(8,0)
\path(6,0)(12,0)
\multiput(0,0)(2,0){7}{\path(0,-.1)(0,.1)}
\thicklines
\path(0,1)(10,1)
\path(2,2)(10,2)
\path(4,3)(12,3)
\put(0,1){\circle*{.15}}
\put(2,2){\circle*{.15}}
\multiput(4,1)(0,2){2}{\circle*{.15}}
\put(6,2){\circle*{.15}}
\put(8,3){\circle*{.15}}
\multiput(10,1)(0,1){3}{\circle*{.15}}
\put(12,3){\circle*{.15}}
\put(-.1,.2){$0$}
\put(1.9,.2){$a$}
\put(3.8,.2){$2a$}
\put(5.8,.2){$3a$}
\put(7.7,.2){$b-a$}
\put(9.9,.2){$b$}
\put(11.5,.2){$2b-a$}
\put(-.3,1.1){$\C$}
\put(1.7,2.1){$\B$}
\put(3.7,1.1){$\A$}
\put(3.7,3.1){$\A$}
\put(5.7,2.1){$\A$}
\put(7.7,3.1){$\B$}
\put(8.1,3.1){$\C$}
\put(9.7,1.1){$\A$}
\put(9.7,2.1){$\A$}
\put(9.7,3.1){$\B$}
\put(10.1,1.1){$\C$}
\put(10.1,2.1){$\B$}
\put(10.1,3.1){$\C$}
\put(11.7,3.1){$\A$}
\end{picture}
\caption{\label{qqdr5} $\s$-Reduction of the QQD scheme, $\s\in R^5$}
\end{center}
\end{figure}
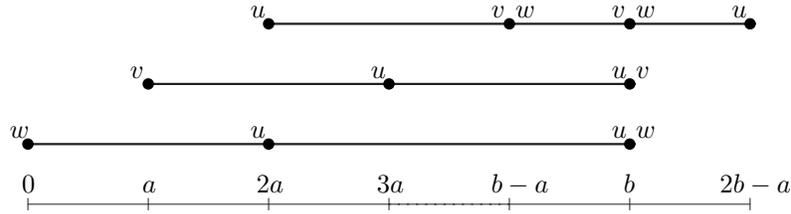

In $R^5$ the values of $\B$ and $\C$ at $n=b$ can be determined independently, using the
second and third equation, respectively. Next the values of $\A$ at $n=2b-a$ can be determined using
the first equation.

\newpage
\section{A non-invertible reduction of the QQD scheme}\label{appendix6}
Using the same notation as in section \ref{ss3}, consider initial values $\{x_0,x_1,x_2,x_3,x_4,$
$y_0,y_1,z_0,z_1\}$ for the $(2,-1)$ reduction of the QQD scheme. We use equation (\ref{eq:QQDQ1}) to
determine $x_5$, equation (\ref{eq:QQDQ2}) to determine $z_2$ and equation (\ref{eq:QQDD}) to determine $y_2$.
Thus we find the nine dimensional mapping
\be
\begin{split}
x_i &\mapsto x_{i+1}, \quad 0\leq i < 5, \\
x_4 &\mapsto x_5 = \frac{y_0x_0}{y_0}, \\
y_0 &\mapsto y_1, \\
y_1 &\mapsto x_1+y_0-z_0+z_2-x_5, \\
z_0 &\mapsto z_1, \\
z_1 &\mapsto z_2 = \frac{x_5}{x_2}z_1.
\end{split}
\ee
for which the staircase method does yield five functionally independent integrals. However, the mapping is not invertible
and so the initial value problem is not well-posed. This example illustrates that in order to prove the well-posedness
of an initial value problem one has to show the initial values can be updated in both directions.

\end{document}